\documentclass[aps,nofootinbib,floatfix,showpacs,amsmath,amssymb]{revtex4-1}

\usepackage{graphicx}
\usepackage{dcolumn}
\usepackage{bm}
\usepackage{url}
\usepackage{epsfig}
\usepackage{latexsym}
\usepackage{amsmath}
\usepackage{amsfonts}   
\usepackage{amssymb}    

\newcommand{\newc}{\newcommand}

\newcommand{\be}{\begin{equation}}
\newcommand{\ee}{\end{equation}}
\newcommand{\bea}{\begin{eqnarray}}
\newcommand{\eea}{\end{eqnarray}}
\newcommand{\sigman}{\sigma_{N}} 
\newcommand{\mdm}{m_{\rm DM}} 

\newc{\mh}{m_h}
\newc{\hone}{h_1} \newc{\mhone}{m_{\hone}}
\newc{\htwo}{h_2} \newc{\mhtwo}{m_{\htwo}}
\newc{\hthree}{h_3} \newc{\mhthree}{m_{\hthree}}
\newc{\aone}{a_1} \newc{\maone}{m_{\aone}}
\newc{\atwo}{a_2} \newc{\matwo}{m_{\atwo}}

\newc{\msew}{m_{S}}
\newc{\mssq}{m_{S}^{2}}
\newc{\alambda}{A_{\lamda}} \newc{\akappa}{A_{\kappa}}

\newc\mueff{\mu_{\text{eff}}} 

\def\eq#1{Eq.~(\ref{#1})}

\def\e3{$\epsilon_3$}

\def\ch2{$\chi^2$}

\def\co#1{{\ifmmode{\cal O}_{#1}\else${\cal O}_{#1}$\fi}}

\newdimen\unit
\def\point#1 #2 #3{\vbox to0pt{\kern-#2\unit
  \hbox{\kern#1\unit#3}\vss}
 \nointerlineskip}

\newcommand{\mev}{\mbox{ MeV}}
\newcommand{\gev}{\mbox{ GeV}}
\newcommand{\tev}{\mbox{ TeV}}

\newcommand{\alphaemmz}{\alpha_{\text{em}}(M_Z)^{\overline{MS}}}
\newcommand{\alphas}{\alpha_s(M_Z)^{\overline{MS}}}

\newcount\hour
\newcount\minute
\newtoks\amorpm
\hour=\time\divide\hour by60 \minute=\time{\multiply\hour by60
\global\advance\minute by- \hour}
\edef\standardtime{{\ifnum\hour<12 \global\amorpm={am}%
    \else\global\amorpm={pm}\advance\hour by-12 \fi
    \ifnum\hour=0 \hour=12 \fi
    \number\hour:\ifnum\minute<100\fi\number\minute\the\amorpm}}
\edef\militarytime{\number\hour:\ifnum\minute<100\fi\number\minute}
\def\bold#1{\setbox0=\hbox{$#1$}%
     \kern-.025em\copy0\kern-\wd0
     \kern.05em\copy0\kern-\wd0
     \kern-.025em\raise.0433em\box0 }

\newc\eg{{\rm {e.g.}}}  \newc\etal{{\rm {et al.}}} \newc\ie{{\rm i.e.}}
\newc\etc{{\rm {etc}}}
\newcommand\lsim{\mathrel{\rlap{\lower4pt\hbox{\hskip1pt$\sim$}}
    \raise1pt\hbox{$<$}}}
\newcommand\gsim{\mathrel{\rlap{\lower4pt\hbox{\hskip1pt$\sim$}}
    \raise1pt\hbox{$>$}}}

\newc{\mhalf}{m_{1/2}}      \newc{\mzero}{m_0}
\newc{\tanb}{\tan\beta}
\newc{\azero}{A_0}
\newc{\at}{A_t} \newc{\ab}{A_b} \newc{\atau}{A_\tau}
\newc{\bmu}{B\mu}           \newc{\sgn}{{\rm sgn}}
\newc{\mone}{M_1}           \newc{\mtwo}{M_2}     \newc{\mthree}{M_3}

 \newc{\hu}{H_u}       \newc{\hd}{H_d}
 \newc{\mhu}{m_{H_u}}       \newc{\mhd}{m_{H_d}}
 \newc{\mhuew}{m^{\ast}_{H_u}}       \newc{\mhdew}{m^{\ast}_{H_d}}
 \newc{\mhuewsq}{m^{\ast\, 2}_{H_u}}       \newc{\mhdewsq}{m^{\ast\, 2}_{H_d}}
 \newc{\mhuast}{m^{\ast}_{H_u}}       \newc{\mhdast}{m^{\ast}_{H_d}}

\newc{\charone}{\chi_1^\pm} \newc{\mcharone}{m_{\chi_1^\pm}}

\newc{\hl}{h}               \newc{\mhl}{m_{\hl}}   \newc{\gammahl}{\Gamma_{\hl}}
\newc{\hh}{H}               \newc{\mhh}{m_{\hh}}   \newc{\gammahh}{\Gamma_{\hh}}
\newc{\ha}{A}               \newc{\mha}{m_{\ha}}   \newc{\gammaha}{\Gamma_{\ha}}
\newc{\hpm}{H^{\pm}}        \newc{\mhpm}{m_{\hpm}} \newc{\gammahpm}{\Gamma_{\hpm}}
\newc{\hp}{H^{+}} 
\newc{\mhp}{m_{\hp}} 
\newc{\mhm}{m_{H^-}}
\newc{\xt}{X_{t}}           \newc{\xb}{X_{b}}

\newc{\qzero}{Q_0}          \newc{\qstop}{Q_{\widetilde t}}
\newc{\amu}{a_{\mu}}        \newc{\amususy}{a_{\mu}^{\text{SUSY}}}
\newc{\amuexpt}{a_{\mu}^{\text{expt}}}        \newc{\amusm}{a_{\mu}^{\text{SM}}}
\newc{\deltaamususy}{\delta a_{\mu}^{\text{SUSY}}}
\newc\gmtwo{(g-2)_{\mu}}
\newc\deltagmtwo{\delta (g-2)_{\mu}} 
\newc\deltaamu{\Delta a_{\mu}}
\newc{\msbar}{\overline{MS}} \newc{\drbar}{\overline{DR}}
\newc{\yt}{h_t} \newc{\yb}{h_b} \newc{\ytau}{h_{\tau}}

\newc{\mtop}{m_t}               \newc{\mtpole}{M_t}
\newc{\mtaupole}{m_{\tau}^{\text{pole}}}
\newc{\mtmtsmmsbar}{m_t(m_t)^{\msbar}_{{\text{SM}}}}
\newc{\mtmtsmdrbar}{m_t(m_t)^{\drbar}_{{\text{SM}}}}
\newc{\mtmtmssmdrbar}{m_t(m_t)^{\drbar}_{{\text{SUSY}}}}

\newc{\mbmbmsbar}{m_b(m_b)^{\msbar} }

\newc{\mbmbsmmsbar}{m_b(m_b)^{\msbar}_{{\text{SM}}}}
\newc{\mbmzsmmsbar}{m_b(\mz)^{\msbar}_{{\text{SM}}}}
\newc{\mbmzsmdrbar}{m_b(\mz)^{\drbar}_{{\text{SM}}}}
\newc{\mbmzmssmdrbar}{m_b(\mz)^{\drbar}_{{\text{SUSY}}}}

\newc{\mtaumzsmmsbar}{m_{\tau}(\mz)^{\msbar}_{{\text{SM}}}}
\newc{\mtaumzsmdrbar}{m_{\tau}(\mz)^{\drbar}_{{\text{SM}}}}
\newc{\mtaumzmssmdrbar}{m_{\tau}(\mz)^{\drbar}_{{\text{SUSY}}}}

\newc{\mgut}{M_{\rm GUT}}
\newc{\mplanck}{M_{\rm P}}      \newc{\mpl}{M_{\text{Pl}}}
\newc{\msusy}{M_{\rm SUSY}}      \newc{\ms}{M_{\text{S}}}
\newc{\jxf}{J({\xf})}
\newc{\jxfexact}{J_{\rm exact}({\xf})}  \newc{\jxfexp}{J_{\rm exp}({\xf})}
\newc{\VEV}[1]{\langle #1 \rangle}
\newc{\xf}{x_f}
\newc\vrel{v_{\rm rel}}
\newcommand\mchi{m_{\chi}}              
\newc\sell{{\widetilde e}_L}      \newc\msell{m_{\sell}}
\newc\selr{{\widetilde e}_R}      \newc\mselr{m_{\selr}}
\newc\snue{{\widetilde \nu}_e}      \newc\msnue{m_{\snue}}
\newc\snutau{{\widetilde \nu}_\tau}      \newc\msnutau{m_{\snutau}}
\newc\supl{{\widetilde u}_L}      \newc\msupl{m_{\supl}}
\newc\supr{{\widetilde u}_R}      \newc\msupr{m_{\supr}}
\newc\sdl{{\widetilde d}_L}      \newc\msdl{m_{\sdl}}
\newc\sdr{{\widetilde d}_R}      \newc\msdr{m_{\sdr}}

\newcommand\sleptonL{{\widetilde l}_L} \newcommand\msleptonL{m_{\sleptonL}}
\newcommand\sleptonR{{\widetilde l}_R} \newcommand\msleptonR{m_{\sleptonR}}
 
\newcommand\selectronL{{\widetilde e}_L} \newcommand\mselectronL{m_{\selectronL}}
\newcommand\selectronR{{\widetilde e}_R} \newcommand\mselectronR{m_{\selectronR}}

\newcommand\smuonL{{\widetilde \mu}_L} \newcommand\msmuonL{m_{\smuonL}}
\newcommand\smuonR{{\widetilde \mu}_R} \newcommand\msmuonR{m_{\smuonR}}

\newcommand\stauL{{\widetilde \tau}_L} \newcommand\mstauL{m_{\stauL}}
\newcommand\stauR{{\widetilde \tau}_R} 
\newcommand\stauone{{\widetilde \tau}_1}   \newcommand\mstauone{m_{\stauone}}

\newcommand\aparam{\widetilde A}

\newc\sfermion{\tilde f}  \newc\msfermion{m_{\sfermion}}
\newc\cmeter{{\rm cm}} \newc\meter{{\rm m}} \newc\kmeter{{\rm km}}
\newc\second{{\rm sec}}

\newc\sr{{\rm sr}}

\newc{\gstar}{g_\ast}           \newc{\gsstar}{g_{s\ast}}
\newc{\geff}{g_{\rm eff}}
\newcommand\mz{m_{Z}}

\newc{\sthw}{\sin\theta_W}              \newc{\cthw}{\cos\theta_W}
\newc{\bino}{\widetilde B}              \newc{\wino}{\widetilde W_30}
\newc{\higgsinob}{{\widetilde H}^0_b}   \newc{\higgsinot}{{\widetilde H}^0_t}
\newc{\abund}{\Omega h^2}
\newc{\abundchi}{\Omega_\chi h^2}
\newc{\abundcdm}{\Omega_{\text{CDM}} h^2}
\newc{\omegam}{\Omega_{M}}       \newc{\abundm}{\Omega_{M} h^2}
\newc{\omegab}{\Omega_{b}}       \newc{\abundb}{\Omega_{b} h^2}
\newc{\omegacdm}{\Omega_{CDM}}
\newc{\omegatot}{\Omega_{TOT}}
\newc{\rhocrit}{\rho_{crit}}
\newc{\rhochi}{\rho_{\chi}}
\newcommand\pb{\,\mbox{pb}} 

\newc\pc{\,\mbox{pc}} \newc\kpc{\,\mbox{kpc}}
\newc\mpc{\,\mbox{Mpc}} \newc\gpc{\,\mbox{Gpc}}

\newc\BR{BR}

\newc\bsgamma{b\rightarrow s \gamma }
\newc\bxsgamma{\overline{B}\rightarrow X_{s}\gamma}
\newc\brbsgamma{\BR(\overline{B}\rightarrow X_s\gamma)}

\newcommand\brbsmumu{\BR(\overline{B}_s\to\mu^+\mu^-)}

\newcommand\butaunu{B_u\to\tau\nu_\tau}
\newcommand\brbutaunu{\BR(\butaunu)}

\newc{\beq}{\begin{equation}}
\newc{\eeq}{\end{equation}}

\newcommand\vs{{\it {vs.}}}

\newc\stoponetwo{{\widetilde t}_{1,2}}
\newc\sbotonetwo{{\widetilde b}_{1,2}}
\newc\stauonetwo{{\widetilde \tau}_{1,2}}

\newc{\sigsip}{\sigma^{\rm SI}_{p}} \newc{\sigsin}{\sigma^{\rm SI}_{n}}
\newc{\sigsiN}{\sigma^{\rm SI}_{N}}
\newc{\sigsdp}{\sigma^{\rm SD}_{p}} \newc{\sigsdn}{\sigma^{\rm SD}_{n}}
\newc{\sigsiA}{\sigma^{\rm SI}_{A}}

\newc{\pbar}{\bar{p}}

\newc{\egamma}{E_{\gamma}}
\newc{\flux}[1]{\Phi_{#1}}
\newc{\dfluxde}[1]{\frac{d\Phi_{#1}}{d E_{#1}}}

\newc{\fluxg}{\Phi_{\gamma}}
\newc{\dfluxgde}{\frac{d\Phi_{\gamma}}{d\egamma}}
\newc{\dfluxgdetext}{ d\Phi_{\gamma} / d\egamma}

\newc{\eplus}{e^+}
\newc{\epos}{E_{\eplus}}
\newc{\eps}{\varepsilon}

\newc{\npos}{n_{\eplus}} \newc{\Npos}{N_{\eplus}}
\newc{\dnposde}{\frac{d n_{\eplus}}{d\epos}}
\newc{\dnposdeps}{\frac{d n_{\eplus}}{d\eps\phantom{_{\eplus}}}}
\newc{\dnposdepstext}{ d n_{\eplus} / d\eps}
\newc{\fluxpos}{\Phi_{\eplus}}  \newc{\fluxelec}{\Phi_{e^{-}}}
\newc{\dfluxposde}{\frac{d\Phi_{\eplus}}{d\epos}}
\newc{\dfluxposdetext}{ d\Phi_{\eplus} / d\epos}

\newc{\nfwc}{{\text{NFW+ac}}} \newc{\moorec}{{\text{Moore+ac}}}

\newc{\chisq}{\chi^2}  \newc{\chisqred}{\chi^2_{\text{red}}}

\newc\xilim{\xi_{\rm lim}}
\newc\tlim{t_{\rm lim}} 
\newc\zetalim{\zeta_{\rm lim}}

\newc\zetah{\zeta_h}
\newc{\relprobone}[1]{p({#1} \vert d)}
\newc{\relprobtwo}[2]{p({#1},{#2} \vert d)}

\long\def\begincomment#1\endcomment{%
        \begingroup\sf\baselineskip12pt#1\endgroup}

\newcommand{\squishlist}{
   \begin{list}{$\bullet$}
    { \setlength{\itemsep}{0pt}      \setlength{\parsep}{3pt}
      \setlength{\topsep}{3pt}       \setlength{\partopsep}{0pt}
      \setlength{\leftmargin}{1.em} \setlength{\labelwidth}{1em}
      \setlength{\labelsep}{0.5em} } }
\newcommand{\squishend}{
    \end{list}  }
        
\def    \be            {\begin{equation}}
\def    \ee            {\end{equation}}
\def    \bea           {\begin{eqnarray}}
\def    \eea           {\end{eqnarray}}

\def\jhep{J. High Energy Phys. }
\def\ds@jhep{\def\@journal{jhep}}

\def\ds@jhep{\def\@journal{jcap}}

\def\ds@plb{\def\@journal{plb}}

\def\ds@prep{\def\@journal{prep}}

\def\ds@cpc{\def\@journal{cpc}}

\def\ds@ijmpa{\def\@journal{ijmpa}}

\def\ds@hepph{\def\@journal{hepph}}

\def\ds@app{\def\@journal{app}}

\begin{document}

\title{Is light neutralino as dark matter still viable?}

\author{Daniel T. Cumberbatch$^{1}$
  \footnote{D.Cumberbatch@sheffield.ac.uk},~Daniel~E.~L\'opez-Fogliani$^{1,2}$~\footnote{daniel.lopez@th.u-psud.fr},~Leszek~Roszkowski$^{1,3}$~\footnote{L.Roszkowski@sheffield.ac.uk},  
  Roberto Ruiz de Austri$^4$ \footnote{rruiz@ific.uv.es}, Yue-Lin
  S. Tsai$^{1,3}$\footnote{Sming.Tsai@fuw.edu.pl}}
\affiliation{
$^1$Department of Physics and Astronomy, The University of Sheffield, 
Sheffield S3 7RH, England\\
$^2$Laboratoire de Physique Th\'eorique, Universit\'e Paris-Sud, F-91405 Orsay, France\\
$^3$The Andrzej Soltan Institute for Nuclear Studies, Warsaw, Poland\\
$^4$Instituto de F\'isica Corpuscular, IFIC-UV/CSIC, Valencia, Spain\\
}

\begin{abstract}
  Motivated by the recent re-confirmation by CoGENT of the low-energy
  excess of events observed last year, and the recent improved limits
  from the XENON-100 experiment that are in contention with the CoGENT
  data, we re-examine the low mass neutralino region of the Minimal
  Supersymmetric Standard Model and of the Next-to-Minimal
  Supersymmetric Standard Model, both without assuming gaugino mass
  unification. We make several focused scans for each model,
  determining conservative constraints on input
  parameters. We then determine how these constraints are made
  increasingly stringent as we re-invoke our experimental constraints
  involving the dark matter relic abundance, collider constraints from
  LEP and the Tevatron, and then from flavour physics, as a series of
  successive $2\sigma$ hard cuts.  We find that for both models, when
  all relevant constraints are applied in this fashion, we 
  do not generate neutralino LSPs that possess a spin-independent 
  scattering cross section in excess of $10^{-5}\pb$ and a mass
  $7\gev\lsim \mchi \lsim 9\gev$
  that is necessary in order to explain the CoGENT observations.
\end{abstract}


\maketitle


\section{Introduction}
\label{sec:intro}

Recent data from several direct detection (DD) dark matter (DM)
experiments have significantly heightened interests in the search and
identification of the elusive DM, that is predicted to dominate the
matter component of the Universe, in the form of weakly-interacting
massive particles (WIMPs). These include the observations made by the
CoGENT experiment~\cite{cogent}, a $p$-type point contact Germanium
detector, whose enhanced sensitivity, due to an unprecedented combination
of target mass and reduced electronic noise, revealed a possible excess of
low energy events~\cite{Aalseth:2010vx} that has recently been re-affirmed in 
June 2011~\cite{Aalseth:2011wp}. If the exponential trend in
these low recoil energy events is fully attributed to the elastic
scattering of WIMP DM, then this signal is consistent with limits on
the DM mass in the range $7\gev\lesssim\mdm\lesssim9\gev$ and
nucleon scattering cross section  in the range $4\times10^{-41}\,{\rm
  cm}^{-2}\lesssim\sigman\lesssim1.5\times10^{-40}\,{\rm cm}^{-2}$,
respectively~\cite{Aalseth:2010vx, Aalseth:2011wp}. 
Remarkably, the CoGENT results favour a WIMP mass
consistent with that indicated by the annual modulation count
rate\footnote{In their June 2011 paper, CoGENT also
claim to observe evidence of an annual modulation in their data that is
consistent with the DAMA/LIBRA results~\cite{Aalseth:2011wp}. }
measured by DAMA/LIBRA~\cite{DAMA, Bernabei:2010mq}, assuming elastic
scattering and that $\sigman$ is spin-independent (SI)~\cite{Petriello:2008jj, 
Chang:2008xa, Fairbairn:2008gz, Savage:2008er, Savage:2009mk}.

However, the majority of the region in $\mdm$\,-\,$\sigman$ parameter
space favoured by CoGENT appears to be in contention with recent
results from direct detection experiments, in particular, those
recently reported by XENON-100~\cite{XENON}, whose 2011 data consist
of non-observations that appear to completely exclude the region
favoured by CoGENT~\cite{Aprile:2011hi}. The CoGENT and DAMA/LIBRA results are
also in contention with the recent observation of two ``anomalous''
events by CDMS-II~\cite{cdms, Ahmed:2009zw}, although the
discrepancies here are much less extensive than with XENON-100, and such
events are not statistically significant enough to be able to claim a
detection of DM (see, e.g.,~\cite{Fitzpatrick:2010em}). Some authors
have previously concluded that the extent of the discrepancies
between the results from CoGENT and DAMA and those from the other DD experiments may be
partially reconciled by, for example, assuming a different proportion
of elastic scattering events at DAMA/LIBRA that are channeled,
adopting DM halo models (specifically pertaining to the escape
velocity and velocity distribution of WIMPs), assuming a lower
scintillation efficiency for Xenon, or assuming different couplings
between WIMPs and either protons or neutrons (see e.g.,~\cite{Fitzpatrick:2010em} 
and references therein). 

Motivated by past experimental findings, several authors have
investigated the parameter space of various particle physics models
capable of generating light WIMP DM candidates that may potentially
explain the CoGENT and DAMA/LIBRA observations (see, e.g.,~\cite{Draper:2010ew, 
Das:2010ww, Vasquez:2010ru, Fornengo:2010mk,
  Cao:2011re}). However, many such investigations achieve this by
conducting focused scans with finely-tuned parameters that result in
points lying close to the CoGENT/DAMA favoured region, but provide a
very poor fit to astrophysical and collider constraints.

In light of the recent release of the 2011 data from CoGENT and
XENON-100, we follow a similar line and re-examine and directly
compare the general behaviour of the neutralino low mass region of two
popular particle physics models: the Minimal Supersymmetric Standard
Model (MSSM) and the Next-to-Minimal Supersymmetric Standard Model
(NMSSM).  We select to investigate the NMSSM in addition to the
popular MSSM since in the NMSSM we expect that a lower mass lightest
supersymmetric particle (LSP) is more easily allowed, relative to that
say in the MSSM, owing to the weakening of the LEP bounds on the Higgs
mass and couplings to Standard Model (SM) particles due to its
extended Higgs sector.

To conduct our investigation we use the nested sampling (NS) algorithm,
implemented in the Bayesian interface tool \texttt{Multinest}~\cite{multinest}, 
that is incorporated into the \texttt{SuperBayeS}~\cite{superbayes} 
and NMSPEC~\cite{nmssmtool} packages, to perform 
scans focused on the low neutralino mass region when imposing
experimental constraints in the likelihood function. However, unlike
other recent studies, we then also impose the same constraints on the
parameter space via a series of (2$\sigma$) hard cuts on the resulting
scan data. These hard cuts consist of an increasing number of
constraints involving, firstly, those on the cosmological DM relic
abundance, as inferred from cosmic microwave background (CMB) and
galaxy cluster observations, secondly, from direct search limits
inferred from particle colliders, e.g., LEP and the Tevatron, and,
thirdly, from indirect constraints from flavour physics and the
anomalous magnetic moment of the muon $\gmtwo$.  For each set of hard
cuts we highlight the parameter space for each of our models that is
consistent with the limits on the LSP mass and SI nucleon scattering
cross section that are imposed by DD limits.

The structure of this paper is as follows: In
Sec.\,\ref{sec:procedure} we describe the details of the scanning
procedure that we utilise to investigate the low mass sectors of our
two particle physics models.  In Sec.\,\ref{sec:mssm} we present and
discuss the results of our scan of the MSSM obtained when imposing our
three sets of constraints from (i) the DM relic density, (ii) collider
physics (excluding flavour physics) and (iii) indirect limits (flavour
physics and $\gmtwo$).  In Sec.\,\ref{sec:NMSSM} we present the
results of our analogous scans on the NMSSM, and compare these results
with those corresponding to the MSSM.
Finally, in Sec.\,\ref{sec:summary} we summarise our conclusions.\\

\section{Our procedure}
\label{sec:procedure}

We are particularly interested in the light neutralino LSP region
of a few GeV. Since this window is excluded in the usual MSSM by LEP
constraints, we relax the assumption of gaugino mass
unification. We do the same also in our study of the NMSSM, even
though in this model a neutralino LSP can potentially evade LEP
constraints more easily than that in the MSSM, and can therefore be
much lighter, because of the possibility of a sizeable singlino
component. We are particularly interested as to how light the
neutralino LSP in the MSSM and the NMSSM can be and which experimental
constraints dominate in defining the lower limit on its
mass. Therefore, in this study we explore the low neutralino mass
regions of both the MSSM and NMSSM, in each case by performing a
single, focused NS scan over a wide range of input parameters.

\begin{table}[b]
\begin{tabular}{|c|c|c|}
  \hline\hline
  Category & Experimental constraint & Reference\\ \hline\hline
  DM relic density & $\abundchi= 0.1123\pm0.0035\ (\mathrm{expt})\ \pm
  0.1\,\abundchi\ (\mathrm{th}) $ & \cite{Jarosik:2010iu}\\ \hline 
  Flavour physics &  $\brbsgamma/10^{-4} =3.55\pm0.26\
  (\mathrm{expt})\ \pm0.21\ (\mathrm{th})$&  \cite{Barberio:2007cr}
  \\ 
  &  $\brbutaunu/10^{-4} = 1.41\pm0.43\ (\mathrm{expt})\ \pm0.38 (\mathrm{th}) $ & \cite{Eriksson:2008cx}\\
  &  $\brbsmumu < 5.8\times 10^{-8}$ (95$\%$\,C.L.)& \cite{Eriksson:2008cx} \\
  &  $\deltagmtwo/10^{-10} = 29.5\pm8.8\ (\mathrm{expt})\ \pm1\ (\mathrm{th})  $ & \cite{Eriksson:2008cx}\\  
  \hline\hline
  Collider & $\Gamma_{\rm{inv}} = 499\pm1.5\mev$ & \cite{pdg}\\ 
  (MSSM only)       & $\mcharone>94\gev$ (95$\%$\,C.L.)  & \cite{pdg} \\ 
  & $\mhone = g(\mha)$ ($2\sigma$ hard cut only)    &  \cite{Sopczak:2001tk}\\
  & $m_{\tilde{e}_R}>73\gev$ (95$\%$\,C.L.) & \cite{pdg} \\
  & $m_{\tilde{\mu}_R}>94\gev$ (95$\%$\,C.L.)    &  \cite{pdg}\\
  & $\mstauone>81.9\gev$ (95$\%$\,C.L.)    & \cite{pdg}  \\
  & $\mha = f(\mhone)$ ($2\sigma$ hard cut only)  &  \cite{Sopczak:2001tk} \\
  \hline
  Collider & As implemented in NMSSMTools. & \cite{nmssmtool}\\
  (NMSSM only)&&\\
  \hline\hline
  Nuisance & $\mtpole= 171.4\pm 2.1\gev\ (\mathrm{expt})$ &  \cite{pdg} \\
  parameters & $m_b (m_b)^{\overline{MS}} =4.20\pm0.07\gev (\mathrm{expt})$ &  \cite{pdg} \\
  & $\alphas = 0.1176\pm 0.002\ (\mathrm{expt})$ & \cite{pdg}\\
  & $1/\alphaemmz=127.955\pm 0.018 \ (\mathrm{expt})$ (MSSM only)& \cite{pdg}\\
  \hline\hline
\end{tabular}
\caption{{\it Top:} a list of experimental constraints relating to
	DM relic density and flavour physics that are invoked
	in the respective likelihood functions, using a Gaussian 
	approximation, associated with our scans of the MSSM 
	and NMSSM.
	{\it Centre:} a list of collider constraints that are
	invoked into the likelihood function, using a Gaussian 
	approximation~\cite{rrt3}, associated with our scan of the MSSM.
	The exception to this is the constraint on $m_A$ and $m_h$,
	which we approximate in the likelihood via a half-Gaussian
	corresponding to $\mhone>89\,$GeV. Then, later, as we do for all other 
	collider constraints, we invoke the constraints
	given by $f(\mhone)$ and $g(m_A)$ as a $2\sigma$ hard cut on the points 
	selected by our initial scan.
	The collider constraints invoked in our corresponding 
	scan of the NMSSM are invoked as hard cuts
	in the respective likelihood function. 
	The procedure in which collider constraints 
	are invoked in our scan of the NMSSM 
	via the NMSSMTool code is significantly more
	complex than that which we adopt for the MSSM,
	and we refer the reader to~\cite{nmssmtool} for
	further details.	
	{\it Bottom:} a list of the nuisance parameters, and their
	corresponding ranges of values, over which our scans 
	are performed.
	Unless otherwise stated, all ranges 
	of values provided are displayed along with their corresponding
	$1\sigma$ experimental (expt) and, where indicated, theoretical (th) errors.}
\label{table:constraints}
\end{table}  

In Table\,\ref{table:constraints} we summarise the sets of
experimental and theoretical constraints that we apply to our scans
of the MSSM and NMSSM.  In the top part of
Table\,\ref{table:constraints} we list the constraints relating to
DM relic density and flavour physics that we invoke in the respective
likelihood functions, using conventional Gaussian probability density
functions (pdfs), as described in, e.g.,~\cite{rrt3}.  The central part of
Table\,\ref{table:constraints} lists the collider constraints that we
include in the MSSM likelihood function, again using a Gaussian
approximation, associated with our corresponding scan. 
The exception to this is the constraint on the mass of the pseudoscalar
Higgs $m_A$ which takes the form of a non-trivial function 
$f(\mhone)$ (with a corresponding inverse function $g(m_A)$) of the lightest CP-even Higgs 
mass $\mhone$ (and $\tanb$); see, e.g., \cite{Sopczak:2001tk} for further details. 
We approximate this constraint in our initial scan by invoking the
constraint $\mhone>89\,$GeV in the likelihood using a half-Gaussian
approximation (we then later invoke the full functional constraints  
given by $f(\mhone)$ and $g(m_A)$ as a 2$\sigma$ hard cut; see below).
We also invoke all relevant collider constraints in the NMSSM likelihood function during
our respective scan. These constraints, which are specified in the
\texttt{NMSSMTool} code, are applied in our scan only as
  2$\sigma$ hard cuts in the relevant likelihood function due to
computational challenges. Because of the complicated process in which
collider constraints are invoked in the NMSSM we omit them in
Table\,\ref{table:constraints} and instead refer the reader to~\cite{nmssmtool} 
for further details.  Finally, the bottom part of
Table\,\ref{table:constraints} lists the nuisance parameters, and
their corresponding ranges of values, over which we perform our
scans. Unless otherwise stated, all ranges of values provided in
Table\,\ref{table:constraints} are displayed along with their
corresponding $1\sigma$ experimental (expt) and, where indicated,
theoretical (th) errors.  Following our initial NS scans we then
successively apply hard cuts to the results using the respective
$2\sigma$ ranges of specified combinations of the invoked constraints,
where in the case of $m_A$ and $\mhone$ we
invoke the constraints given by $f(\mhone)$ and $g(m_A)$
as a $2\sigma$ hard cut.

An alternative to the above strategy would be to perform three
separate scans over both MSSM and NMSSM input parameters, omitting
subsequent hard cuts, where in each case using a likelihood function
invoked with a successively increasing number of experimental
constraints, i.e., `likelihood(DM relic density)',
`likelihood(DM relic density+collider)', and
`likelihood(DM relic density+collider+flavour)'. Despite the efficient
nature of such scans, this strategy possesses two problems. Firstly,
using these three different likelihood functions to investigate the
low mass neutralino regime can result in posterior pdfs centred upon
very different regions of parameter space. For example, it is quite
possible that the favoured regions resulting from the
`likelihood(DM relic density+collider)' scan will not significantly overlap 
with that from the `likelihood(DM relic density)' scan, making
statistical arguments regarding the overall favoured region
difficult. Secondly, if we include all experimental constraints into
the likelihood function, the resulting $\chi^2$, which, given our
assumptions and priors, is given by $-2\log[{\rm likelihood}]$, can be
written as
\begin{equation}
\chi^2_{\rm tot}=\chi^2_1+\chi^2_2+...,
\label{eq:chisq}
\end{equation}
where $\chi^2_i$ represents the $\chi^2$ statistics associated with a
particular experimental constraint. 
In NS scans, \eq{eq:chisq} allows for, and often results in, points being selected which 
predict values that satisfy some constraints extremely well but 
lie far outside of the usual $1\sigma$ range of others, often used 
in traditional fixed-grid scans, whilst still yielding acceptable 
$\chi^2_{\rm tot}$ values. Moreover, such a method lacks transparency 
as to how the results of its scans correspond with individual 
experimental constraints.

Since in our scans we apply all experimental constraints as Gaussians in the
likelihood functions (with the exception being those on $m_A$ and
$m_h$; see above), and subsequently make $2\sigma$ cuts on the
corresponding results, we are hence able to overcome both of the above
issues.  However, we emphasise that, because of this we will not
provide a statistical interpretation of the point distribution of our
numerical scans and display all our results as scatter plots,
illustrating only the density of points, rather than pdfs.  We use the
NS scanning technique effectively to take advantage of the fact that
it can perform an efficient scan of a multi-dimensional parameter
space in order to obtain a representative sample of points, for
specified ranges of parameters, allowing us to simply observe the
general behaviour of this parameter space, rather than to draw
statistical conclusions from our scans.

\section{Results in the MSSM}
\label{sec:mssm}

In this section, our goal is to map out the mass range of the light
neutralino LSP in the MSSM that is consistent with current estimates
for the present dark matter relic abundance, as well as constraints,
both direct and indirect, from collider experiments.  It is clear that
because existing experimental limits are so stringent, especially
those from LEP regarding the chargino mass, one needs to consider
scenarios that have the potential to evade them in order to explore
the low mass regime in which we are interested. On the other hand, in
order to conduct a numerical scan that is manageable, one needs to
focus primarily upon the parameters which are relevant to the problem
at hand.

Bearing this in mind, we 
assume minimal flavour violation and perform a scan, as described in
Sec.\,\ref{sec:procedure}, over the following MSSM parameters:
\begin{equation}
\mone, \mtwo, \mu, \tanb, \mha, \msleptonL, \msleptonR,
\label{eq:mssmparams}
\end{equation}
where $\mone$ and $\mtwo$ are the soft bino and wino masses repectively, 
$\mu$ is the Higgs/higgsino mass parameter, $\tanb$ is the ratio of the 
vacuum expectation values (VEVs) of the up and down-type neutral Higgs
fields, 
$\mha$ is the mass of the pseudoscalar Higgs boson, whilst 
$\msleptonL$ and $\msleptonR$ are the left and right hand slepton 
soft mass parameters. (For a more detailed description of the MSSM, we
refer the reader to a dedicated review, e.g.,~\cite{Martin:1997ns}.)
We adopt log priors for all mass input parameters, and a 
flat prior for $\tanb$. The prior ranges of the MSSM parameters
over which our scan is performed are provided in 
Table\,\ref{table:MSSMparams}.

We take all left-handed slepton soft mass parameters to be family
degenerate (i.e., $\mselectronL=\msmuonL=\mstauL=\msleptonL$), and
likewise for their right-handed partners.  On the other hand, gluino
and squark masses will not play any significant role in determining
either the relic density $\abundchi$ or the spin-independent
elastic scattering cross section $\sigsip$, and hence we fix them at
$1\tev$.  Likewise, we fix all the trilinear couplings at the
electroweak (EW) scale: $\at=\ab=\atau=\aparam=-0.5\tev$.  In order to
focus our scan towards the low mass region of a few GeV, we adopt a
log prior distribution of the bino mass $\mone$, which
primarily determines the mass, $\mchi$, of the lightest
neutralino. Also we note that allowing for a more generous range of
parameters by expanding the lower limits beyond those displayed would
result in our scan finding a disproportionate number of points in the
region inconsistent with collider constraints. Note that, because we
relax the usual assumption of gaugino mass unification (i.e.,
$\mthree:\mtwo:\mone \simeq 6:2:1$), and take the prior range of
$\mone$ well below those of $\mtwo$ and of $\mu$, in order to more
efficiently generate light, bino-like neutralino LSP that can evade
LEP constraints on the invisible $Z$ width and the chargino mass~\cite{Fitzpatrick:2010em}. 
Another parameter that will play an
important role in our scan is $\tanb$ since it affects both $\mchi$ as
well as the fermion-sfermion couplings of the neutralino LSP,
especially in the low mass regime that we are interested in, and we
take a very wide range of its values in our scans.

\begin{table}[t]
\begin{tabular}{|c|c||c|c|}
\hline
\hline
Parameter & Range & Parameter & Range\\
\hline\hline
bino mass &  $0.1<\mone<40$ & pseudoscalar mass & $85<\mha<600$ \\
wino mass & $90<\mtwo <150$ & slepton-left mass & $70<\msleptonL<3000$ \\
$\mu$ parameter & $90<\mu<150$ & slepton-right mass & $70<\msleptonR<3000$ \\
ratio of Higgs doublet VEVs & $1<\tanb<62$ &&\\
\hline
\hline
\end{tabular}
\caption{The prior ranges of input parameters 
over which we perform our scan of the MSSM. All displayed mass ranges
are given in GeV. We adopt log priors for all mass input parameters, and a flat prior for $\tanb$}
\label{table:MSSMparams}
\end{table} 

\begin{figure}[b]
	\begin{center}
	\includegraphics[width=0.496\linewidth, keepaspectratio]{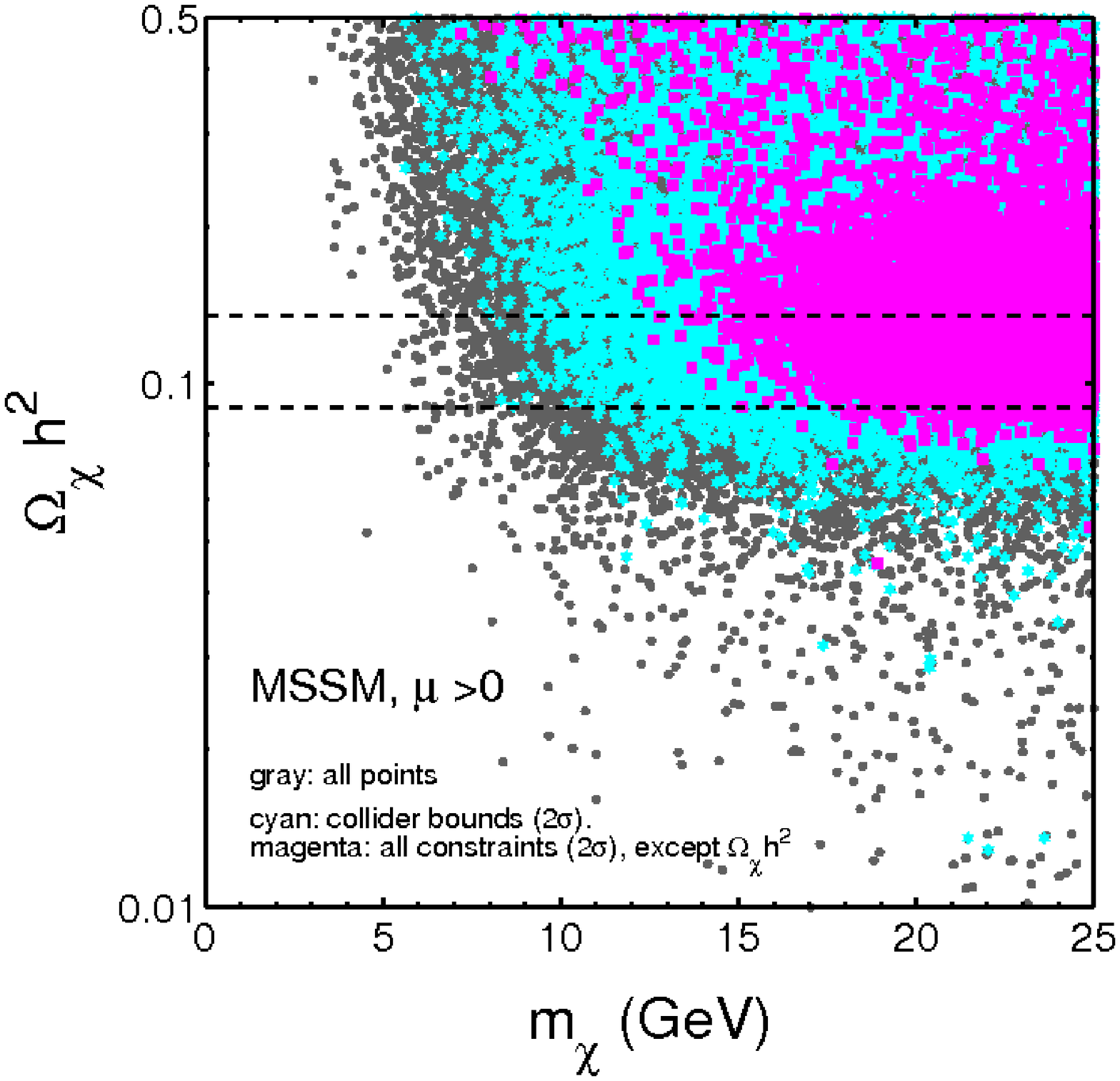}
	\includegraphics[width=0.49\linewidth, keepaspectratio]{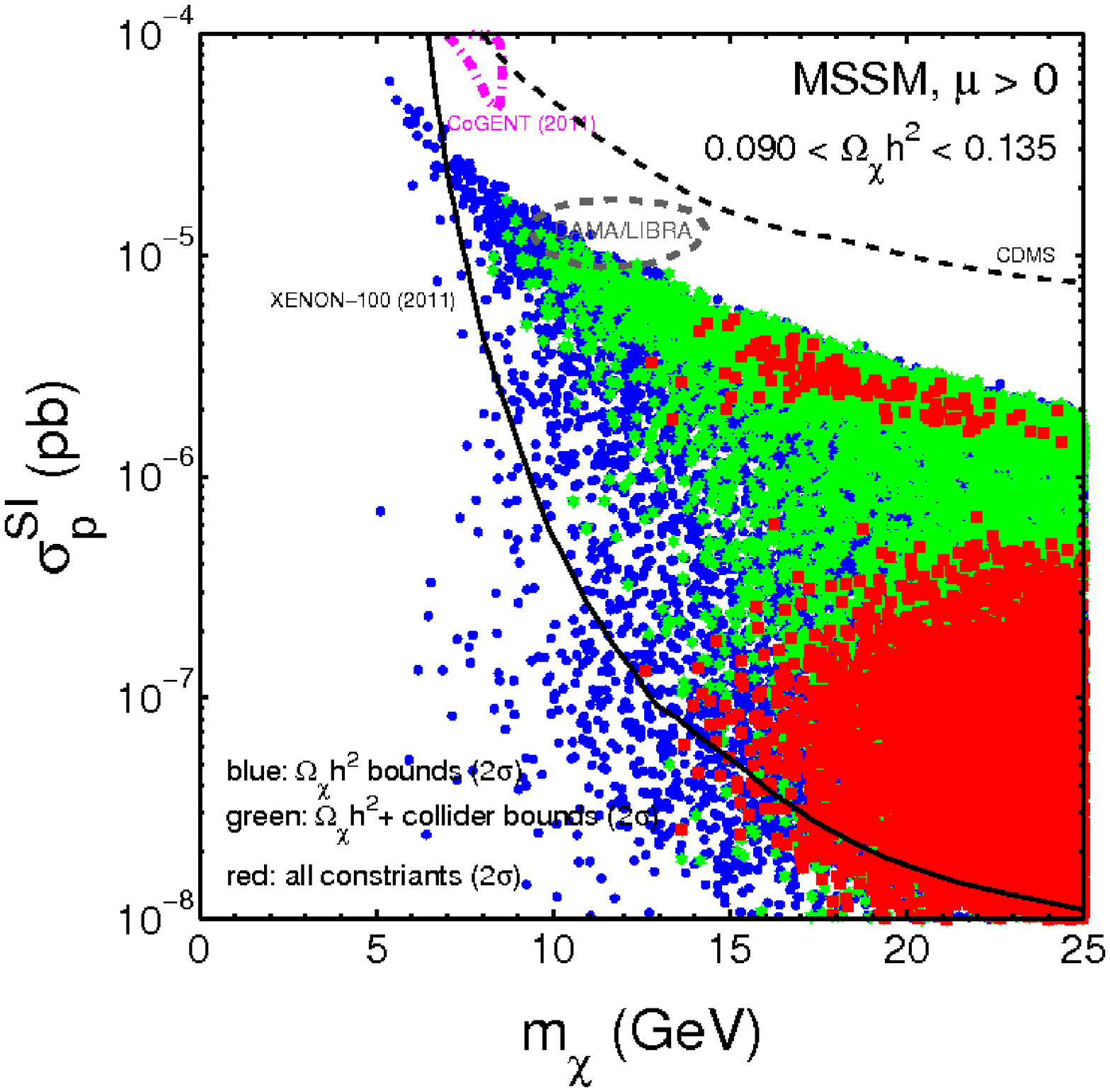}
      \end{center}
      \caption{{\it Left panel:} $\mchi$ \vs\ $\abundchi$ in the MSSM. Grey points:
        all the points resulting from our NS scan, where constraints on
        $\abundchi$, collider limits and flavour physics constraints, 
        as specified in Table\,\protect\ref{table:constraints}, are
        included in the likelihood. Cyan points: points from our scan
        surviving subsequent $2\sigma$ hard cuts using collider
        bounds.  Magenta points: points from our scan surviving
        subsequent $2\sigma$ hard cuts using both collider and flavour
        physics constraints.  We also indicate the $2\sigma$ range of
        relic densities within the experimental best-fit value (black
        dashed curves).  {\it Right panel:} $\mchi$ \vs\ $\sigsip$ in
        the MSSM.  Blue points: configurations in our initial NS
        scan yielding relic densities in the range:
        $0.090<\abundchi<0.135$ (i.e., $2\sigma$ range).  Green points:
        points that survive a $2\sigma$ hard cut using both relic
        density and collider bounds.  Red points: MSSM parameter
        configurations surviving $2\sigma$ hard cuts using all
        constraints listed in Table\,\ref{table:constraints}.  We note
        that the results corresponding to $\abundchi\le\Omega_{\rm
          DM}h^2$ are very similar to those displayed in this
        panel following our hard cuts, with only a handful of
        additional points being found in the region: $\mchi\gsim10
        \gev$, $\sigsip\lsim 2\times10^{-5}\pb$.  We also illustrate
        the regions of parameter space currently favoured by the
        CoGENT (outlined by the magenta dash-dotted curve) and
        DAMA/LIBRA (with ion channelling, outlined by the dashed grey
        curve) experiments, and display the current limits from CDMS-II
        (dashed black curve) and XENON-100 (solid black curve).}
      \label{fig:oh2mchisigmchi_mssm}
\end{figure}

To start with, in the left panel of 
Fig.\,\ref{fig:oh2mchisigmchi_mssm} 
we present the results of our MSSM scan in the $\mchi$\,-\,$\abundchi$
plane. Grey points denote all the points resulting from our NS 
scan, where constraints on $\abundchi$, collider limits and flavour 
physics constraints, 
as specified in Table\,\protect\ref{table:constraints}, are included
in the likelihood. Since we display our results in the 
$\mchi$\,-\,$\abundchi$ plane we of course omit any hard cuts on the relic
abundance, but instead simply illustrate the corresponding $2\sigma$
range (dashed lines).

Some features are clearly visible. Firstly, we observe a great number
of points that predict a relic density that far exceeds the
corresponding $2\sigma$ upper limit, particularly those in the low
$\mchi$ region. However, we also observe that there are points in this
region that predict a relic density within the $2\sigma$ limits
indicated by the black dashed curves. This is predominantly achieved
through efficient $t$-channel annihilations via light stau exchange,
where the $\chi\tau\stauR$ coupling, which scales as $\tan^2\beta$,
and hence, is enhanced at large $\tanb$. In fact, we find no points
with $\tanb<4$ for $\mchi<35\gev$ that predict a relic density within
the experimental $2\sigma$ range.  This result is simply a reflection
of the SUSY neutralino equivalent of the Lee-Weinberg bound on massive
neutrinos~\cite{Lee:1977ua}.  By requiring $\abundchi$ to be within
the $2\sigma$ range, we find $\mchi\gsim5\gev$, which is indeed
consistent with the results previously obtained in~\cite{Fornengo:2010mk}. 
Such low mass points appear to be absent in
Fig.\,3 of~\cite{Vasquez:2010ru}, which finds $\mchi\gtrsim11\gev$,
although, since cuts were made at the $68\%\,$C.L., it is possible that
points possessing smaller $\mchi$ but poorer likelihood were generated
there.

Unfortunately, obtaining points with the desired small values of 
$\mchi$ of a few GeV comes at the expense of potentially grossly 
violating other constraints, as mentioned in 
Sec.\,\ref{sec:procedure}. In order to see this, we impose a hard 
cut of $2\sigma$ on all of the collider constraints listed in 
Table\,\ref{table:constraints}. The points surviving this cut are 
displayed in the left panel of Fig.\,\ref{fig:oh2mchisigmchi_mssm} 
in cyan. We can clearly observe that a sizeable number of the points
found in our scan (i.e. grey points), especially predicting low 
$\mchi$ and/or low $\abundchi$, badly violate collider limits. 
We also find that, because of the strong correlation between the 
lower limit on $\mchi$ and the minimum slepton mass, invoking the 
LEP constraint on $\mselectronR$, $\msmuonR$ and $\mstauone$ are 
responsible for the observed increase in the lowest value of 
$\mchi$ possessed by a point generated from our scan that satisfies 
the $2\sigma$ bounds on $\abundchi$ from roughly $5\gev$ (for the 
grey points in the left panel of 
Fig.\,\ref{fig:oh2mchisigmchi_mssm}) to approximately $7\gev$ when 
LEP constraints are invoked (corresponding to the cyan points in the
left panel of Fig.\,\ref{fig:oh2mchisigmchi_mssm}).
Finally, we impose the remaining constraints from flavour physics 
and the anomalous magnetic moment of the muon, once again by making 
a $2\sigma$ hard cut on the initial set of points selected in our 
NS scan. The surviving points are shown in the left panel of 
Fig.\,\ref{fig:oh2mchisigmchi_mssm} in magenta. 

Following the above sets of hard cuts we can clearly observe that 
whilst the LSP with mass as 
small as $5\gev$ still survives, which may lie within the range of 
WIMP mass favoured by CoGENT and DAMA/LIBRA (with ion 
channelling; see, e.g.,~\cite{Savage:2008er}), only points with 
$\mchi\gsim 13\gev$, which lie outside of the ranges favoured by 
these experiments, predict values of $\abundchi$ consistent
with the corresponding $2\sigma$ constraint.

In the right panel of Fig.\,\ref{fig:oh2mchisigmchi_mssm} we
demonstrate, this time in the $\mchi$\,-\,$\sigsip$ plane, the effect of
applying an increasing number of our constraints.  The blue points
represent the configurations in our initial NS scan that yield a
value of $\abundchi$ within the favoured $2\sigma$ range
$0.090<\abundchi<0.135$. The green points illustrate those of the 
blue points that survive a $2\sigma$ hard cut using collider bounds.
Finally, the red points correspond to those MSSM parameter
configurations that survive hard cuts using all of the constraints 
in Table\,\ref{table:constraints}, imposed at $2\sigma$.  

We note that we also conducted separate scans, identical to those
described above except that the neutralino LSP generated was allowed
to contribute only {\it partially} to the currently inferred DM
abundance by invoking a $2\sigma$ upper bound of $\abundchi<0.135$ in
the likelihood function. The results we found were very similar
to those displayed in the right panel of
Fig.\,\ref{fig:oh2mchisigmchi_mssm}, with only a handful of additional
points being found in the region: $\mchi\gsim10 \gev$,
$\sigsip\lsim2\times10^{-5}\pb$. The robustness of our results with
respect to the constraints on $\abundchi$ is consistent with the
findings of Vasquez {\it et al.}~\cite{Vasquez:2010ru}.  However we
also note that whilst in~\cite{Vasquez:2010ru} it was claimed that 
neutralino mass should obey $\mchi\gtrsim28\gev$ in order to
evade present experimental bounds, determined primarily by the limits
from XENON-100 (see Fig.\,7 of~\cite{Vasquez:2010ru}), from
Fig.\,\ref{fig:oh2mchisigmchi_mssm} (right panel) we find points with 
$\mchi$ as small as $\simeq13\gev$ that
satisfy both the XENON-100 limits as well as our additional 2$\sigma$
hard cuts.

In the MSSM the SI elastic scattering cross section $\sigsip$ is 
primarily determined by the $t$-channel exchange of the heavy Higgs 
scalar $\hh$, with the effective $\chi \hh p$ coupling being 
enhanced by large values of $\tanb$. However, because of LEP bounds 
on the Higgs sector, in the MSSM, one is left with little freedom to
boost $\sigsip$ towards the regions currently favoured by CoGENT 
and DAMA/LIBRA (with ion channelling), as outlined in 
Fig.\,\ref{fig:oh2mchisigmchi_mssm} by the magenta dash-dotted curve
and the dashed grey curve respectively. Moreover, even for the blue 
points in Fig.\,\ref{fig:oh2mchisigmchi_mssm} (right panel), where 
only a $2\sigma$ hard cut on $\abundchi$ is performed, the Higgs mass is not 
entirely unrestricted because of the lower limits of our selected 
prior ranges for MSSM input parameters (see 
Table\,\ref{table:MSSMparams}). We can also see that imposing 
collider bounds has the substantial effect of cutting off the wedge 
of points with $\mchi\lsim8\gev$ possessing the largest SI elastic 
scattering cross sections, $\sigsip\gsim2\times10^{-5}\pb$, which lie
just below the CoGENT favoured region. On the other hand, we note
that collider constraints have a fairly limited effect at larger 
values of $\mchi$.

\begin{figure}[t]
	\begin{center}
 	\includegraphics[width=0.498\linewidth, keepaspectratio]{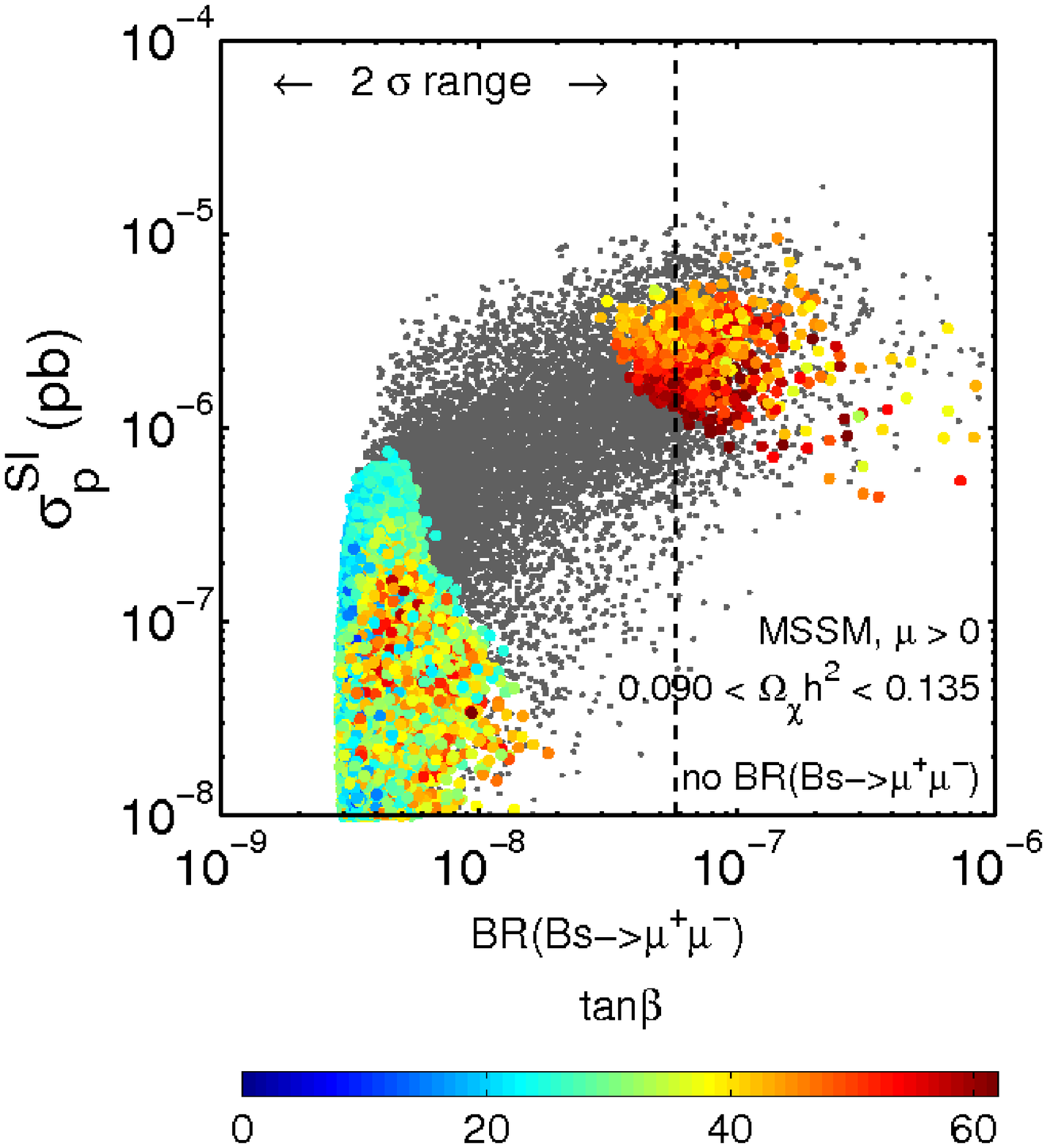}
 	\includegraphics[width=0.485\linewidth, keepaspectratio]{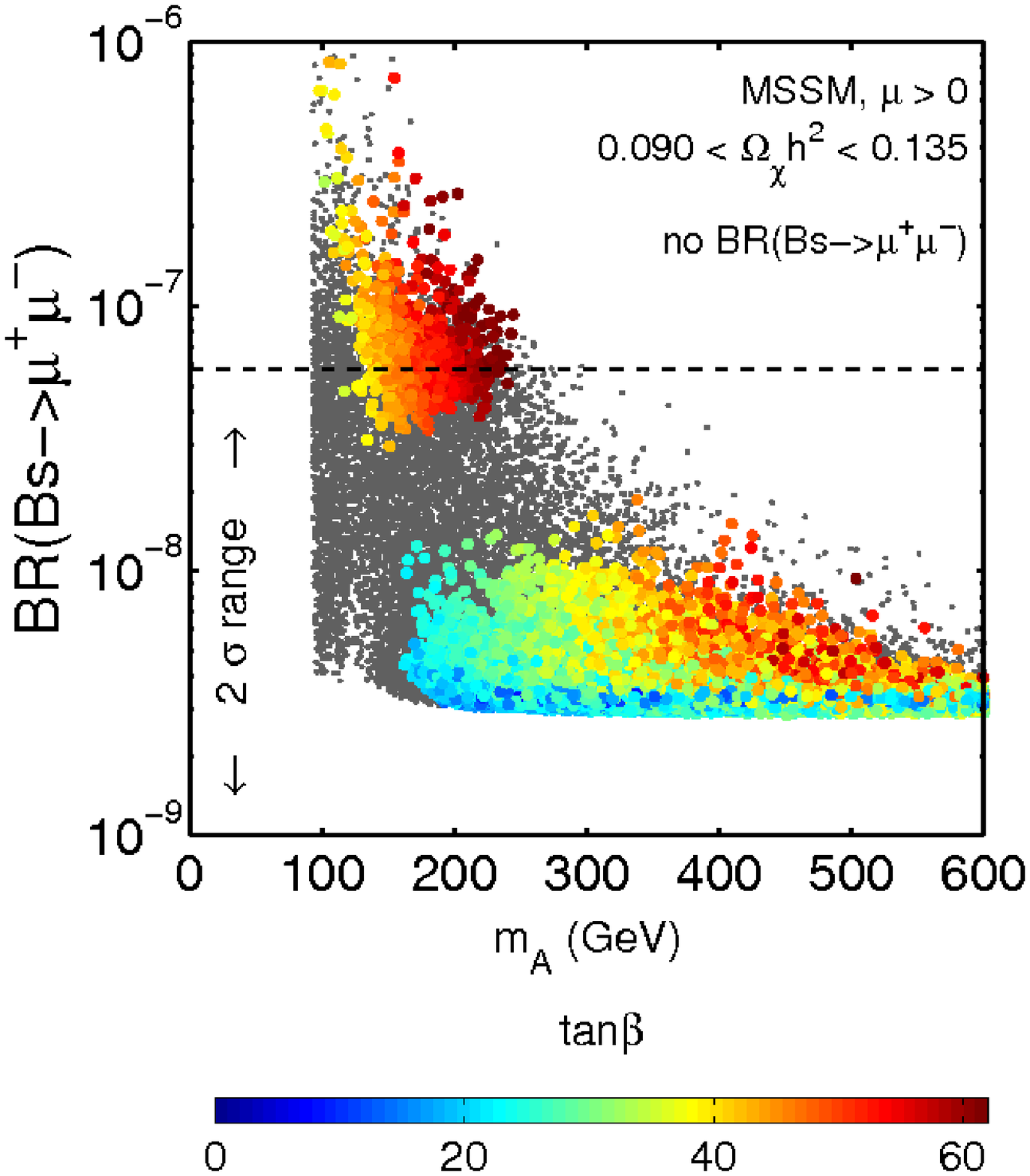}
      \end{center}
         \caption{{\it Left panel:} $\brbsmumu$  \vs\ $\sigsip$ in the
           MSSM. Grey points: points from our initial NS scan 
           which survive subsequent $2\sigma$ hard cuts using 
           both our $\abundchi$ and collider constraints. 
           Coloured points: those of the
           grey points which survive $2\sigma$ hard cuts using 
           our flavour physics constraints, excluding that on 
           $\brbsmumu$.
           The colour scale denotes the value of $\tanb$ 
           associated with each point. We also illustrate the
           Tevatron limit on $\brbsmumu$ (dashed line).
           {\it Right panel:} $\mha\,\vs\ $$\brbsmumu$ in the
           MSSM, using the same colour scale as displayed in the
           left panel.
         }\label{fig:sig_bsmumu_mssm}
\end{figure}

In Fig.\,\ref{fig:oh2mchisigmchi_mssm} (right panel), we observe that
when we impose our $2\sigma$ hard cuts using collider and flavour physics 
constraints a vast number of points in the region: $\mchi\lsim13\gev$,
$\sigsip\gsim 3\times10^{-6}\pb$ are excluded.
Such points are consistent with those found in the fore-mentioned
investigation of the MSSM in~\cite{Fornengo:2010mk} for
$85\gev\lsim m_A\lsim100\gev$ and $\tanb\gtrsim30$. However, as remarked
above, when we invoke flavour physics constraints, particularly those
on $\brbsgamma$ (and also $\brbsmumu$, see below), as $2\sigma$ hard cuts,
such points are clearly excluded. To confirm this, we note that we
performed a separate scan of the MSSM, focused on the range of $m_A$
and $\tanb$ mentioned above and using values of the input parameters
that range over those utilised in~\cite{Fornengo:2010mk}. We found
that the resulting points possessed $\brbsgamma> 4.31\times10^{-4}$
(i.e., $2.3\,\sigma$ result) and hence, from
Table\,\ref{table:constraints}, fail to survive our $2\sigma$ hard
cuts, confirming the results from our larger scans.

Upon further examination, we find that it is fairly easy to generate
points that satisfy the constraints imposed on $\brbsgamma$,
$\brbutaunu$, as well as on $\deltagmtwo$. In particular, the latter
constraint is especially weak since, in order to generate a large
enough SUSY contribution to the relic density, one requires a fairly
light smuon, which, being a free parameter, is easy to produce, which
in turn can give a substantial contribution to $\deltagmtwo$.

On the other hand, an improved upper bound on $\brbsmumu$ from the
Tevatron puts a rather strong limit on the upper range of $\sigsip$
values generated in our scan.  This can be seen in the left panel of
Fig.\,\ref{fig:sig_bsmumu_mssm}, where we plot the values of
$\brbsmumu$ against $\sigsip$ generated by some of the points in our
initial NS scan.  Grey points represent those scan points which
survive subsequent $2\sigma$ hard cuts using both our DM 
and collider constraints, while the coloured points represent those 
of the grey points which survive further $2\sigma$ hard cuts using 
flavour physics constraints, except for that on $\brbsmumu$.  From
Fig.\,\ref{fig:sig_bsmumu_mssm} we also observe the well-known
correlation between $\sigsip$ and $\brbsmumu$, arising from the fact
that $\brbsmumu\propto \tan^6\beta/\mha^4$ and that $\mha\simeq\mhh$
for $\mha\gg100\gev$~\cite{Bobeth:2001sq, Lunghi:2006uf}.  On the
other hand, at smaller $\mha$ ($\sim100\gev$) this correlation 
is relaxed because the approximate mass degeneracy does not hold 
there.

In Fig.\,\ref{fig:sig_bsmumu_mssm} (left panel) we observe two ``islands'' of
points.  The one in the bottom left corner is constituted by points 
possessing low to average values of $\tanb$, which survive the upper
limit on $\brbsmumu$ from the Tevatron, but produce values of
$\sigsip\lsim10^{-6}\pb$, much smaller than those favoured by CoGENT
or DAMA/LIBRA (with ion channelling).  On the other hand, points 
constituting the upper right-hand island generate values of $\sigsip$ up to 
$\sim10^{-5}\pb$, that include the handful of red points located in 
the region: $10^{-6}\lesssim\sigsip\lsim 6\times10^{-6}\pb$, $\mchi\gsim 13\gev$, 
but are mostly in conflict with the upper limit on $\brbsmumu$.  
Such points possess large values of $\tanb$ and also small values of 
$\mha$, reaching very close to the corresponding LEP limits for the 
MSSM, worsening their overall likelihood~\cite{pdg,Sopczak:2001tk}. 
This can be seen from the right panel of Fig.\,\ref{fig:sig_bsmumu_mssm}, where we plot 
the values of $\mha$ against the values of $\brbsmumu$ possessed by the same 
points as displayed in the left panel. A more thorough
treatment of Higgs LEP limits by including them in the likelihood 
could possibly produce more points with a larger $\sigsip$ but again 
they would likely correspond to stretching the LEP limits.

From Fig.\,\ref{fig:sig_bsmumu_mssm} it is clear that in the MSSM, the
effect of imposing all the relevant constraints as hard cuts,
especially that on $\brbsmumu$, even at the somewhat relaxed level of
$2\sigma$, is to exclude any of our points located in either of the
regions favoured by CoGENT or DAMA/LIBRA (with ion channelling).
However, as remarked in Sec.\,\ref{sec:intro}, we note that some authors
(see, e.g.,~\cite{Fitzpatrick:2010em} and references therein) have
illustrated how these favoured regions may be shifted slightly when
taking into account certain uncertainties relating to the velocity
distribution of Galactic DM or the detectors themselves.

\section{Results in the NMSSM}
\label{sec:NMSSM}

In this section, we continue our discussion of the low mass neutralino
region,
now turning to our analysis of the NMSSM, starting briefly with a
summary of some basic properties of the NMSSM. For a
more detailed description of the NMSSM we refer the reader to a
dedicated review (see, e.g.,~\cite{Ellwanger:2009dp}).

\subsection{The NMSSM}
\label{subsec:NMSSM_model}

The NMSSM is an extension of the MSSM that provides a 
solution to the so-called $\mu$-problem~\cite{kn84}. Its superpotential differs to 
that of the MSSM in that it contains a new superfield $S$ which is a
singlet under the SM gauge group 
$SU(3)_c \times SU(2)_L \times U(1)_Y$. (We use the same notation for
superfields and their respective spin-0 component fields for 
simplicity.) The part of the NMSSM superpotential involving Higgs 
fields is given by
\begin{equation}
\label{2:Wnmssm}
W = -\epsilon_{ij} \lambda \,S \,H_d^i H_u^j +\frac{1}{3} \kappa S^3\,,
\end{equation}
where $H_d^T=(H_d^0, H_d^-)$, $H_u^T=(H_u^+, H_u^0)$, $i,j$ are
$SU(2)$ indices with $\epsilon_{12}=-1$, while $\lambda$ and $\kappa$
are dimensionless couplings in the extended Higgs sector. 
The superpotential Eq.\,(\ref{2:Wnmssm}) is scale invariant, and the EW 
scale will only appear through the corresponding soft SUSY breaking 
terms in the soft Lagrangian, $\mathcal{L}_{\text{soft}}$. Using the 
above notational conventions, the terms in $\mathcal{L}_{\text{soft}}$
associated with Eq.\,(\ref{2:Wnmssm}) are given by

\begin{eqnarray}
\label{2:Vsoft}
-\mathcal{L}_{\text{soft}}&=&m_{H_d}^2 \,H_d^*\,H_d + m_{H_u}^2 \,H_u^* H_u + m_{S}^2 \,S^* S \nonumber \\
&-& \left(\epsilon_{ij} \lambda\, A_\lambda S H_d^i H_u^j +
\frac{1}{3} \kappa \,A_\kappa\,S^3 + {\rm {\it h.c.}} \right),\nonumber \\
\end{eqnarray}
where $m_{H_d}, m_{H_u}$ and $m_{S}$ are the soft breaking Higgs 
masses, with $A_\lambda$ and $A_\kappa$ being trilinear soft-breaking 
terms. Consequently, despite that in this model the usual MSSM bilinear 
$\mu$-term is absent from the superpotential
Eq.\,(\ref{2:Wnmssm}), which only possesses dimensionless trilinear couplings,
when the scalar component of $S$ acquires a 
VEV, $s=\langle S\rangle$, an effective interaction 
$\mu_{\rm eff} H_d H_u$ is generated, with the effective parameter 
$\mu_{\rm eff} = \lambda s$, which is then naturally of the EW scale~\cite{Ellwanger:2009dp}.

In addition to terms from $\mathcal{L}_{\text{soft}}$, the tree-level 
scalar Higgs potential receives the usual $D$ and $F$ term 
contributions:

\begin{align}\label{2:Vfd}
V_D = & \, \,\frac{g_1^2+g_2^2}{8} \left( |H_d|^2 - |H_u|^2 \right)^2 +
\frac{g_2^2}{2} |H_d^\dagger H_u|^2 \, , \nonumber \\
V_F = & \, \,|\lambda|^2
\left( |H_d|^2 |S|^2 + |H_u|^2 |S|^2 + |\epsilon_{ij} H_d^i H_u^j|^2 \right)
+ |\kappa|^2 |S|^4
\nonumber \\
&
-\left( \epsilon_{ij} \lambda \kappa^* H_d^{i} H_u^{j}S^{*2} + {\rm {\it h.c.}}
\right) \,,
\end{align}
where $g_1$ is the gauge coupling of the electroweak SU(2)$_L$
and $g_2$ is the gauge coupling of the U(1)$_Y$ of the SM. 
Using the minimization conditions derived for the Higgs VEVs,
one can re-express the soft breaking Higgs masses in terms of 
$\lambda$, $\kappa$, $A_\lambda$, $A_\kappa$,
$v_d=\langle H_d^0 \rangle$, $v_u=\langle H_u^0 \rangle$, and $s$:
 \begin{align}
m_{H_d}^2 = & -\lambda^2 \left( s^2 + v^2\sin^2\beta \right)
- \frac{1}{2} M_Z^2 \cos 2\beta
+\lambda s \tan \beta \left(\kappa s +A_\lambda \right) \,,\nonumber\\
m_{H_u}^2 = & -\lambda^2 \left( s^2 + v^2\cos^2\beta \right)
+\frac{1}{2} M_Z^2 \cos 2\beta
+\lambda s \cot \beta \left(\kappa s +A_\lambda \right) \,,\nonumber\\
m_{S}^2 = & -\lambda^2 v^2 - 2\kappa^2 s^2 + \lambda \kappa v^2
\sin 2\beta + \frac{\lambda A_\lambda v^2}{2s} \sin 2\beta -
\kappa A_\kappa s\,,
\end{align}
\noindent where $v^2=v_1^2+v_2^2=2M_W^2/g_2^2$, and $M_W$ is the mass 
of the W-boson.

Consequently, at the EW scale, the free parameters in the Higgs 
sector (at tree level) are:
$\lambda$, $\kappa$, $m_{H_1}^2$, $m_{H_2}^2$,  $m_{S}^2$, $A_\lambda$ 
and $A_\kappa$.  Using the fore-mentioned minimization conditions of 
the Higgs potential, one can eliminate the soft Higgs masses in 
favour of $M_Z, \tan \beta$ and $\mu$, resulting in $\lambda, \,
\kappa,\, \tan \beta,\, \mu,\, A_\lambda, \, A_\kappa$ as a set of independent parameters.
In our scans we will actually use a somewhat different set of
parameters, including the soft scalar masses as well as the soft
gaugino masses, $M_1, M_2$ and $M_3$ which are all free parameters at the EW
scale; these parameters will all be listed in Sec.\,\ref{subsec:NMSSM_results}.

\subsection{Results}
\label{subsec:NMSSM_results}

In this subsection we present the results of our scan of the NMSSM which were obtained as 
described in Sec.\,\ref{sec:procedure}. In Sec.\,\ref{sec:mssm}, we 
observed that, in the MSSM, even without imposing gaugino 
unification, the constraints on the Higgs sector from LEP and the 
Tevatron were significant in reducing the number of scan points 
possessing small values of $\mchi$ that also yielded a relic density
within $2\sigma$ of the experimental best-fit value.

Analogous to our scan over MSSM parameters, in order to evade LEP
bounds on the chargino mass for our scan of the NMSSM, we again relax
the assumption of universality amongst the gaugino masses: $\mone$,
$\mtwo$ and $\mthree$ at the unification scale.  Consequently, we
allow the bino mass to be very small without the implication of
yielding an unacceptably light chargino. Further, we must also satisfy
LEP constraints on the invisible $Z$ width. Because of the gaugino
non-universality, the majority of the points selected in our MSSM scan
were bino-dominated with a gaugino fraction $g_f>0.6$. Similarly, the
dominant bino composition of our points selected in our NMSSM scan
helps evading the constraint on $\Gamma_Z$, which is proportional to
the difference between the squares of the higgsino components of the neutralino
LSP~\cite{Fitzpatrick:2010em}.

Alternatively, in the NMSSM, one can potentially evade the collider
constraints with the neutralino LSP that possesses a dominant singlino
contribution. Then bino-dominated neutralinos, bino/singlino mixed
neutralinos and singlino-like neutralinos are possible as a light LSP~\cite{Gunion:2005}. 
In these three scenarios, for neutralino masses
below $\sim15\gev$, efficient annihilation is achieved mainly via a
light Higgs resonance exchange~\cite{Belanger:2005kh}.  Therefore, in order to obtain an
acceptable range of the relic density for a very light neutralino, one
needs in general very light Higgses with the mass of the lightest
Higgs scalar $\mhone$ tuned to be close to $2\mchi$. This implies that, in
order to evade LEP bounds, the lightest Higgs must be singlet-like.
In order to generate this kind of  Higgs boson without driving down the
mass of the SM Higgs below the LEP bound of $\simeq114.4\gev$~\cite{Sopczak:2001tk}, 
one needs to decouple the singlet and doublet
Higgs components (i.e., obtain the limit
$\mathcal{M}_{S,13}^2, \mathcal{M}_{S,23}^2 \rightarrow 0$ for the mass
matrix $\mathcal{M}_S$ of CP-even Higgs states (see,
e.g., Eq.\,(2.8) of~\cite{Cerdeno:2004xw})).

As discussed in~\cite{Ellwanger:2006rm},
there are two possible ways to achieve this. One is to adopt the 
limit $\lambda\rightarrow0$. Alternatively, one can enforce the 
relation:
\begin{equation}
A_{\lambda}=\frac{2}{\sin2\beta}\mueff-2\kappa s.
\label{eq:NMSSM_A_lambda}
\end{equation}
Adopting Eq.\,(\ref{eq:NMSSM_A_lambda}) exactly
yields a mass for the lightest CP-even singlet Higgs scalar given by
\begin{equation}
m_{h_1}^{2}=\lambda^{2}A_{\lambda}\frac{v_1v_2}{\lambda s}+\kappa s(A_{\kappa}+4\kappa s)\,,
\label{eq:NMSSM_m_hone}
\end{equation}
which is equal to $M_{S,33}^2$ (once again, see, e.g., Eq.\,(2.8) of~\cite{Cerdeno:2004xw}). 
Such solutions were found in the study
conducted by B\'{e}langer {\it et al.}~\cite{Belanger:2005kh}.  From
Eq.\,(\ref{eq:NMSSM_m_hone}), we can see that there are new types of
solutions present in the NMSSM that are not obtainable in the MSSM. In
the limit $\lambda\rightarrow0$, the mass of the CP-odd neutral scalar
Higgs (i.e., the pseudoscalar Higgs) is given by
\begin{equation}
m_{\rm
  pseudoscalar}^{2}=\lambda(2\kappa+
\frac{A_{\lambda}}{2s})v^{2}\sin2\beta-3\kappa A_{\kappa}s\rightarrow-3\kappa A_{\kappa}s, 
\label{eq:NMSSM_m_pseudoscalar}
\end{equation}
which is equal to $\mathcal{M}_{P,22}^2$, where $\mathcal{M}_P$ is
the mass matrix of CP-odd Higgs states, and where, in the
limit $\lambda\rightarrow0$, $\mathcal{M}_{P,12}\rightarrow0$ and
hence, due to the zero mass terms of the Goldstone boson,
$\mathcal{M_P}$ is diagonal (see, e.g., Eq.~(2.10) of~\cite{Cerdeno:2004xw}). 
From Eq.\,(\ref{eq:NMSSM_m_pseudoscalar}) we
can see that a negative value of $\kappa A_{\kappa}s$ is necessary in
order to obtain $m^2_{\rm pseudoscalar}>0$ and, consequently, avoid
tachyons \footnote{Notice that without lost of generality $\lambda$
  can be chosen  to be positive~\cite{Cerdeno:2004xw} and that with this convention
  the sign of $s$ and the sign of $ \mueff$ are the same.}.  For small
$\lambda$ it will be difficult to achieve a resonance annihilation via
a singlet Higgs when the neutralino is singlino-like due to tendency
for the masses of the neutralino LSP and the Higgs to be
similar. Hence, in this case, a bino-dominated neutralino LSP will be
favoured.

\begin{table}[t]
\begin{tabular}{|c|c||c|c|}
\hline
\hline
Parameter & Range & Parameter & Range \\
\hline\hline
bino mass &  $0.1 < \mone < 30$ & CP-odd neutral scalar Higgs mass & $85 < m_A < 600$ \\
wino mass & $90 < \mtwo < 500$ & slepton-left mass & $70 < \msleptonL < 3000$ \\
$\mueff$ parameter & $90 < \mueff< 500$ & slepton-right mass & $70 < \msleptonR < 3000$ \\
ratio of Higgs doublet s & $2<\tanb<65$ & trilinear terms & $|A_\kappa| < 100$,  $|\tilde{A}| < 4000$\\
Higgs sector coupling & $10^{-4} < \lambda < 0.5$ & Higgs sector coupling & $10^{-4} < \kappa < 0.5$ \\
\hline
\hline
\end{tabular}
\caption{The prior ranges of input parameters over which we perform
  our scan of the NMSSM. All displayed mass ranges are given in
  GeV. We adopt log priors for all input parameters except $\tanb$, for which we use a flat prior.}
\label{table:NMSSMparam}
\end{table} 

\begin{figure}[b]
	\begin{center}
	\includegraphics[width=0.495\linewidth, keepaspectratio]{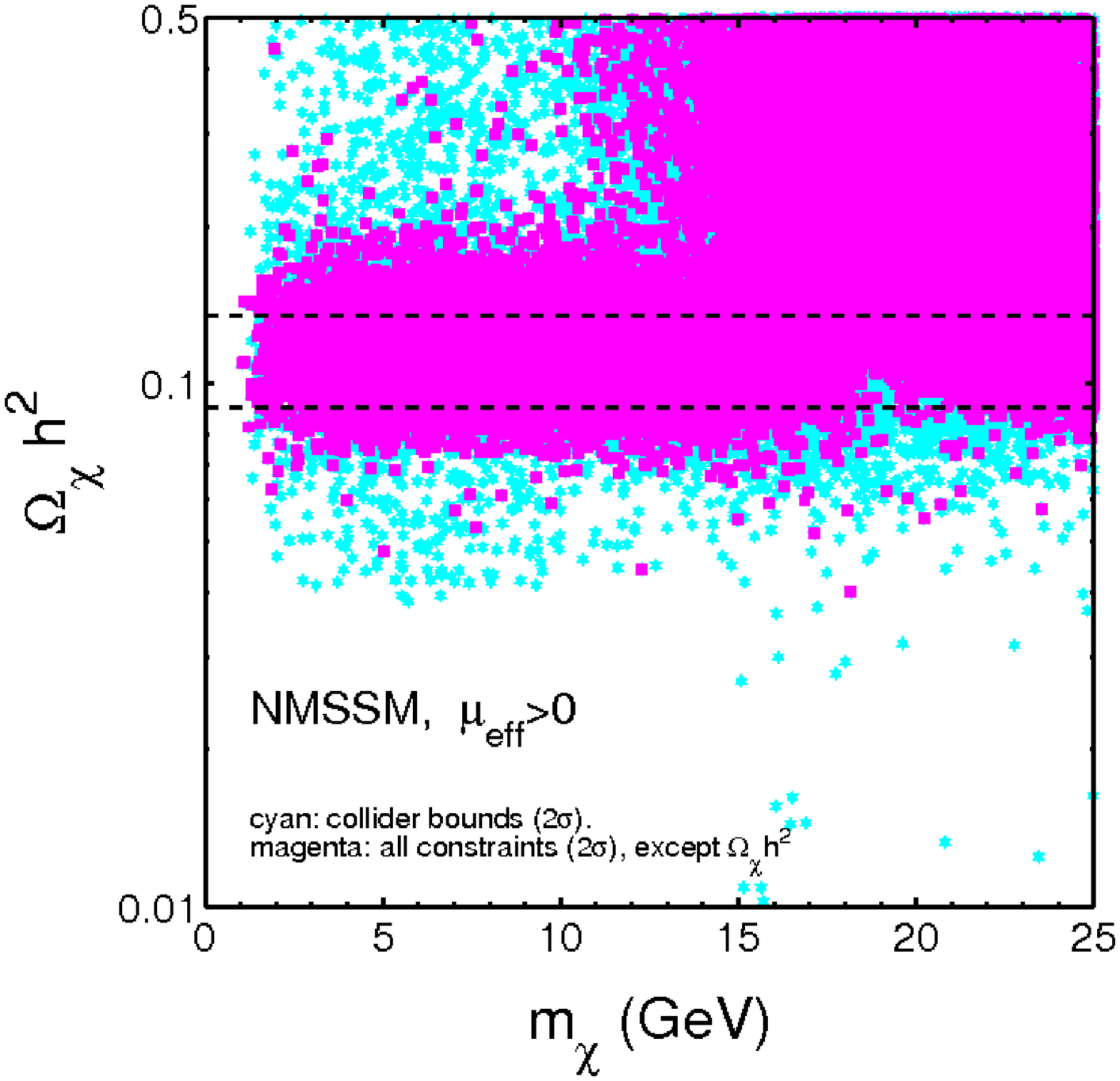}
	\includegraphics[width=0.486\linewidth, keepaspectratio]{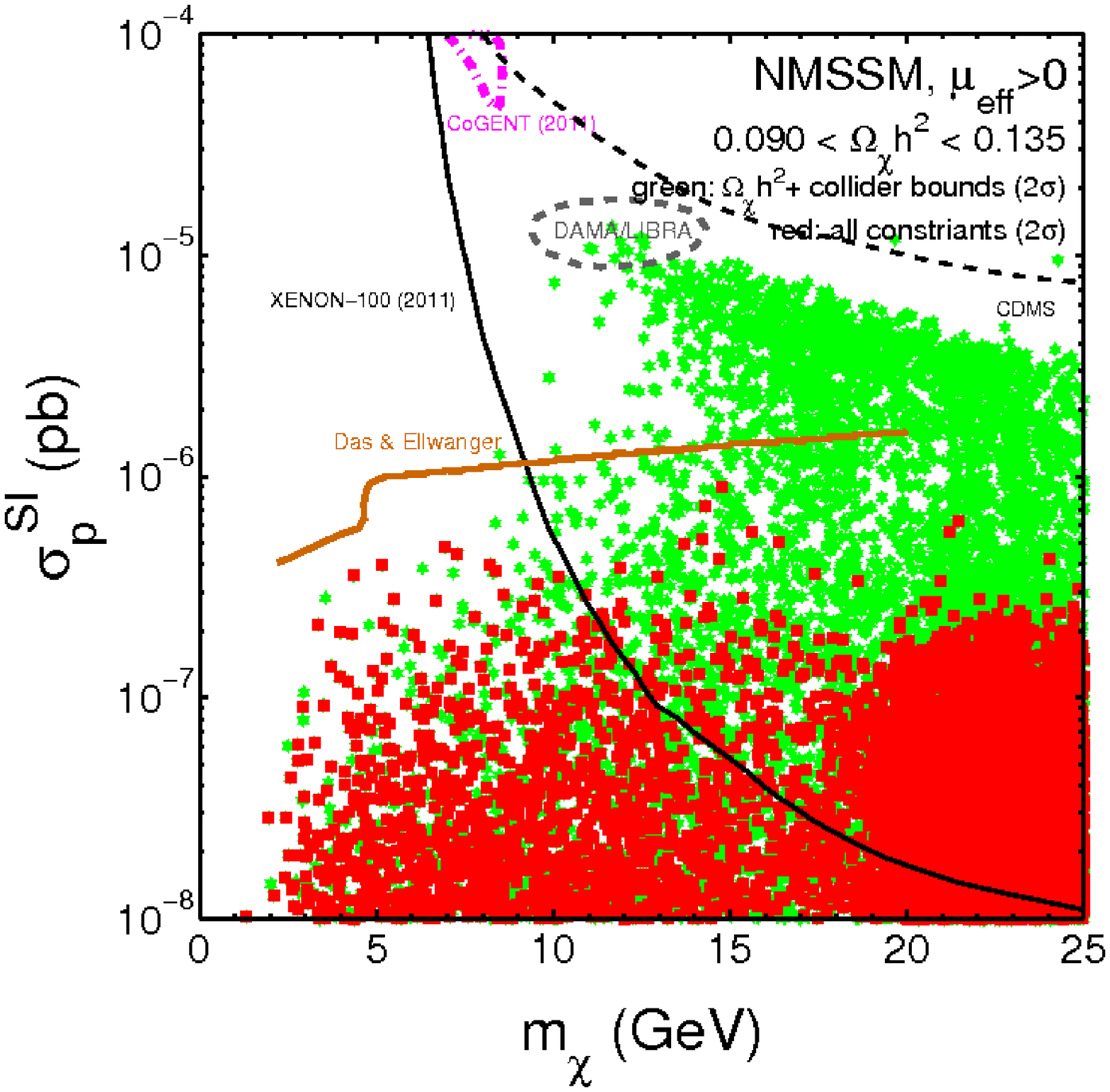}
      \end{center}
      \caption{{\it Left panel:} $\mchi$ \vs\ $\abundchi$ in the
        NMSSM. Cyan points: all the points resulting from our NS
        scan, where constraints on $\abundchi$ and flavour physics
        bounds are included in the likelihood as Gaussians, and
        collider constraints are included in the likelihood as hard
        cuts.  All relevant constraints are specified in
        Table\,\ref{table:constraints}.  Magenta points: points from
        our initial scan surviving subsequent $2\sigma$ hard cuts
        using flavour physics constraints.  We also indicate the
        $2\sigma$ range of relic densities within the experimental
        best-fit value (black dashed curves).  {\it Right panel:}
        $\mchi$ \vs\ $\sigsip$ in the NMSSM.  Green points: those of
        the cyan points in the left panel which survive a subsequent
        $2\sigma$ hard cut on $\abundchi$.  Red points: those of the
        magenta points in the left panel that survive a subsequent
        $2\sigma$ hard cut on $\abundchi$.  We also illustrate the
        regions of parameter space currently favoured by CoGENT
        (outlined by the magenta dash-dotted curve), DAMA/LIBRA (with
        ion channelling, outlined by the dashed grey curve), and
        illustrate the current limits from CDMS-II (dashed black
        curve) and XENON-100 (solid black curve). We also display the
        estimated upper bound on $\sigsip$ in the NMSSM according to
        Das\,\&\,Ellwanger~\cite{Das:2010ww} when using default values
        of strange quark content in nucleons (orange curve).  }
\label{fig:oh2mchisigmchi_nmssm}
\end{figure}

Conversely, for larger $\lambda$, it is possible to generate nearly
pure singlet-like Higgs states by fine-tuning $A_{\lambda}$ according to
Eq.\,(\ref{eq:NMSSM_A_lambda}). These solutions allow for
singlino-like neutralinos that annihilate via a singlet-like Higgs
that are much more efficient than those corresponding annihilations
involving a bino-like LSP.  Since the constraints from LEP on the
chargino mass ultimately yields a lower bound on $\mu_{\rm
  eff}=\lambda s$, and from Eq.\,(\ref{eq:NMSSM_A_lambda}), a
hierarchy is established between $\lambda$ and $\kappa$, where
$\kappa\ll\lambda$ for $\lambda\gg0$, in such a way as to result in a
small value of $\mchi\sim2\kappa s$.
Alternatively, for large $\lambda$, one may avoid this hierarchy
as well as generate a light, pure, singlet-like CP-even or CP-odd Higgs by finely-tuning
$A_\kappa$ to drive down the mass given by Eq.\,(\ref{eq:NMSSM_m_hone}) or
Eq.\,(\ref{eq:NMSSM_m_pseudoscalar}) respectively. Such fine-tuning allows for 
LSPs similar to those found in~\cite{Cao:2011re}.
As a result of these two fine-tuning scenarios, the solutions in the regime $\lambda>0.1$ 
are generally finely-tuned, resulting in mixed bino-singlino LSPs.

Taking into account all our constraints (i.e., DM relic density,
collider and flavour physics), one can determine that a value of
$\lambda\sim 10^{-2}$ favours a bino-dominated neutralino LSP. For
larger values of $\lambda \sim 10^{-1}$ neutralinos are a mixture
of both bino and singlino.  For this reason, the selection of either a
log or flat prior for $\lambda$ can make a huge difference
to numerical results.  
We highlight the caveat that all such solutions are fine-tuned since 
$2m_{\chi}\approx m_{h_1}$ or  $2m_{\chi}\approx m_{a_1}$, 
where $m_{a_1}$ is the mass of the lightest doublet-like pseudoscalar Higgs
(whereas the heavier doublet-like pseudoscalar mass is denoted by $m_{a_2}$), 
however for $\lambda\gtrsim10^{-1}$ we find that a further 
numerical fine-tuning, this time involving $A_\lambda$, is necessary
in order to generate a singlet Higgs pure enough to escape collider constraints. 
One should also
mention that in this scenario collider constraints on the heavier,
MSSM-like, CP-even Higgs $h_2$ become extremely important, favouring small
values for $\tanb\sim3$~\cite{Belanger:2005kh}.

\begin{figure}[t]
	\begin{center}
	\includegraphics[width=0.49\linewidth, keepaspectratio]{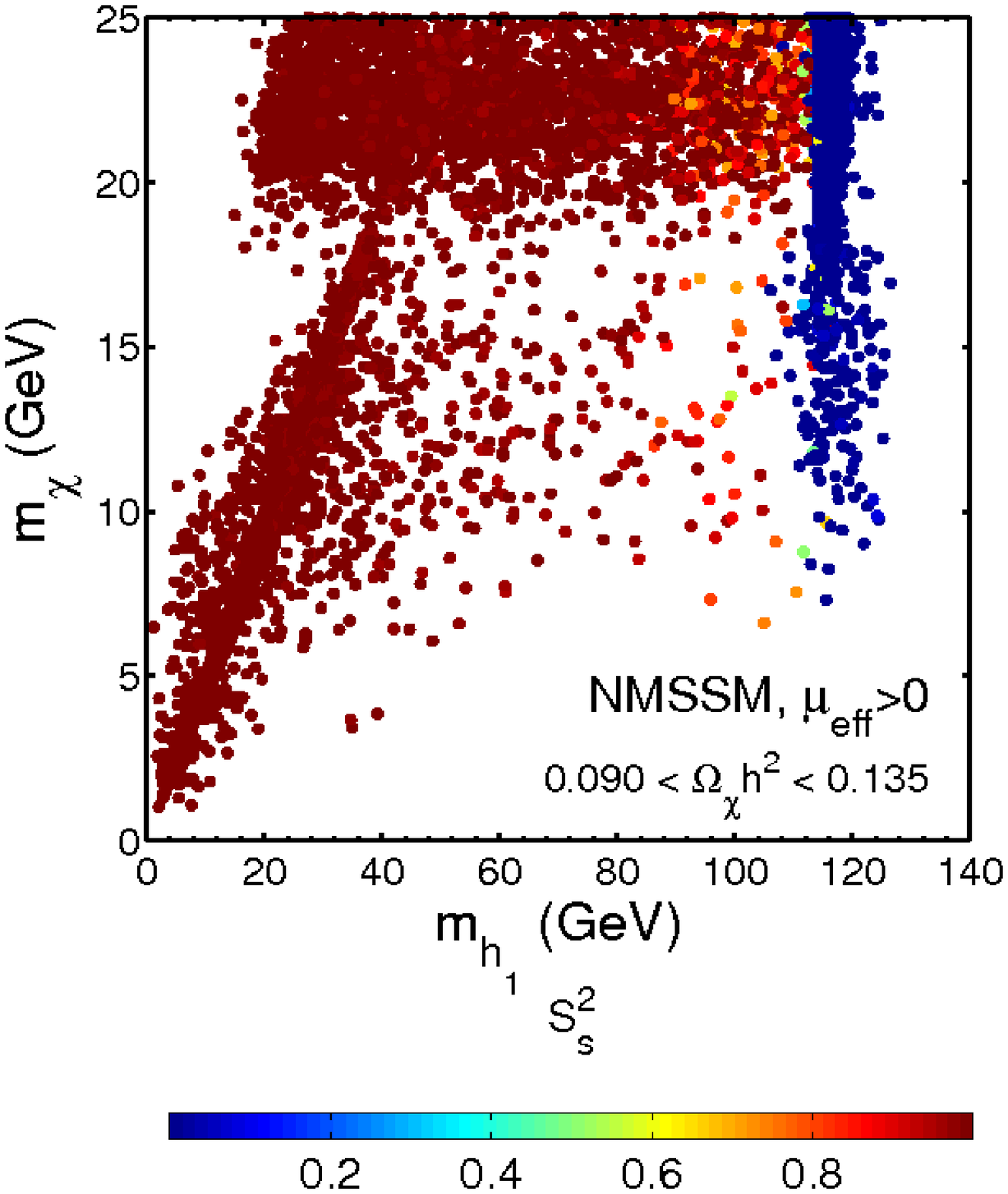}
	\includegraphics[width=0.495\linewidth, keepaspectratio]{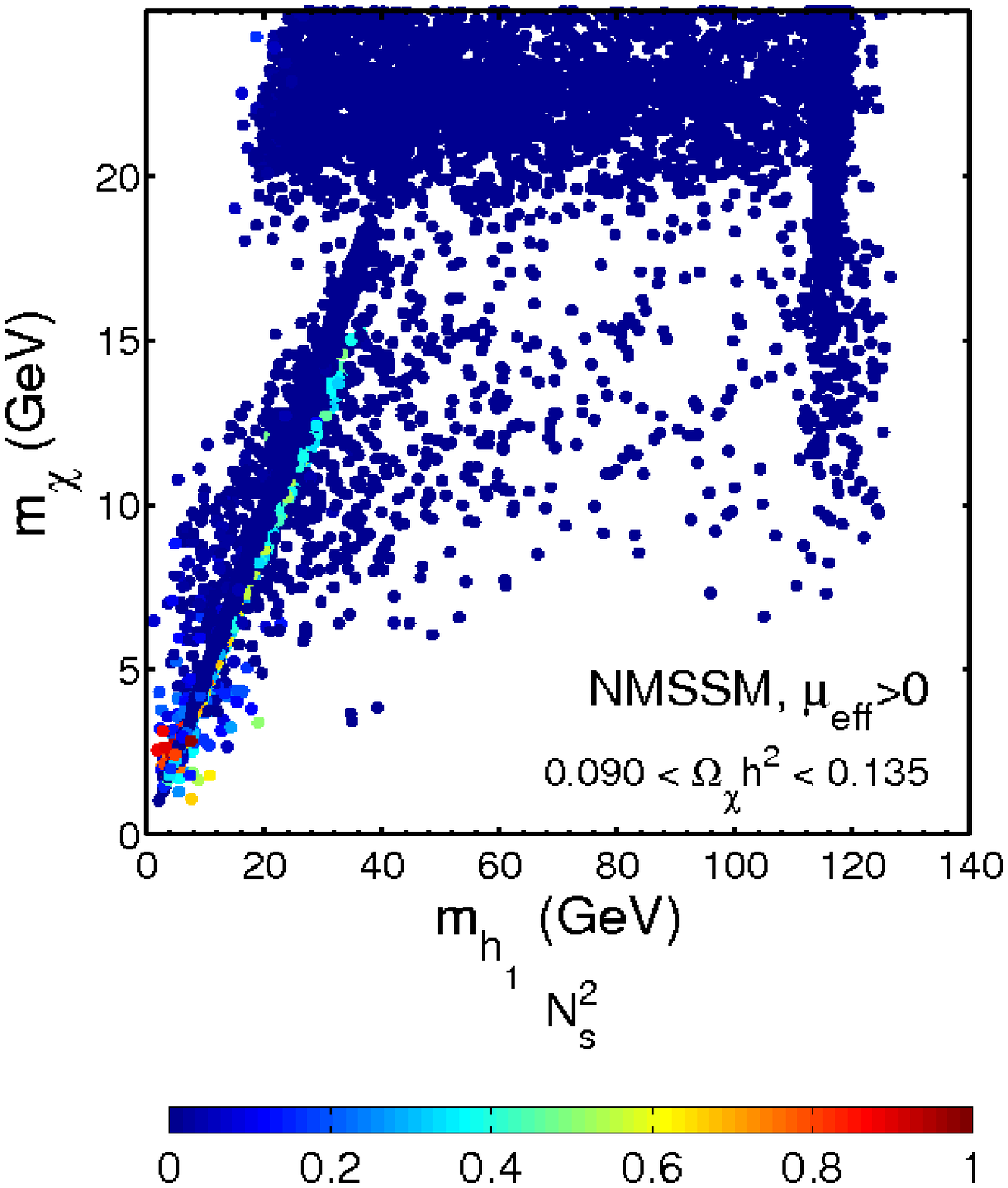}
      \end{center}
      \caption{{\it Left panel:} $m_{h_1}$ \vs\ $\mchi$ in the NMSSM,
      for the points from our initial NS scan that survive 
      subsequent $2\sigma$ hard cuts on $\abundchi$ and
      when using constraints from colliders (including flavour physics).
      The colour scale illustrates the singlet fraction ${\rm S_{\rm S}^2}$
      of the lightest Higgs.
      {\it Right panel:} Identical to the left panel except now the colour scale indicates the 
        singlino fraction of each respective neutralino LSP.  
      }
\label{fig:mh1mchi_nmssm}
\end{figure}

Bearing in mind the above, we performed our scan of the NMSSM, using
the NMSPEC code~\cite{nmssmtool} and imposing all relevant constraints
in the likelihood as described in Sec.\,\ref{sec:procedure}, taken
over the following input parameters:
\begin{equation}
\mone, \mtwo, \mueff, m_{A}, \msleptonL, \msleptonR, A_{\kappa}, \tilde{A}, \tanb, \lambda, \kappa.
\label{eq:NMSSMparam}
\end{equation}

\noindent where $m_A$ denotes the running mass of the lighter CP-odd
Higgs-doublet, as defined in the NMSPEC code~\cite{nmssmtool} 
(see Table\,\ref{table:NMSSMparam}), and the
other parameters in Eq.\,(\ref{eq:NMSSMparam}) have already been 
defined above\footnote{Note that in the NMSPEC code, when $m_A$ 
is chosen as input, $A_\lambda$ is treated as output.}.  where $m_A$ 
is an approximative value for $m_{a_1}$.
Once again, we adopt log priors for
all input parameters except $\tanb$, for which we use a flat prior,
when conducting our scans. 
In particular, we utilise a log prior for
$\lambda$ and $\kappa$ so that we may explore the general behaviour 
of low mass LSPs in the $\lambda$\,-\,$\kappa$ plane, rather
than adopt flat priors and focus on finely-tuned solutions possessing $\lambda\gg0$,
as in, e.g.,~\cite{Belanger:2005kh}.
Also, for the same reasons as, and in order to be consistent, with our scan
of the MSSM, we assume all slepton soft mass parameters as well as
their right-handed partners to be degenerate and fix the gluino and
all squark masses at $1\tev$.  The prior ranges of the NMSSM
parameters Eq.\,(\ref{eq:NMSSMparam}) over which our scan is performed
are provided in Table\,\ref{table:NMSSMparam}. In analogy to our MSSM
scan, we restrict the bino mass to the range $0.1\gev < \mone <
30\gev$ and the higgsino mass parameter to the range $90\gev< \mueff <
150\gev$, in order to focus on light, bino-like neutralino LSP and
evade collider constraints on the Higgs sector, as discussed above.

\begin{figure}[t]
	\begin{center}
 	\includegraphics[width=0.487\linewidth, keepaspectratio]{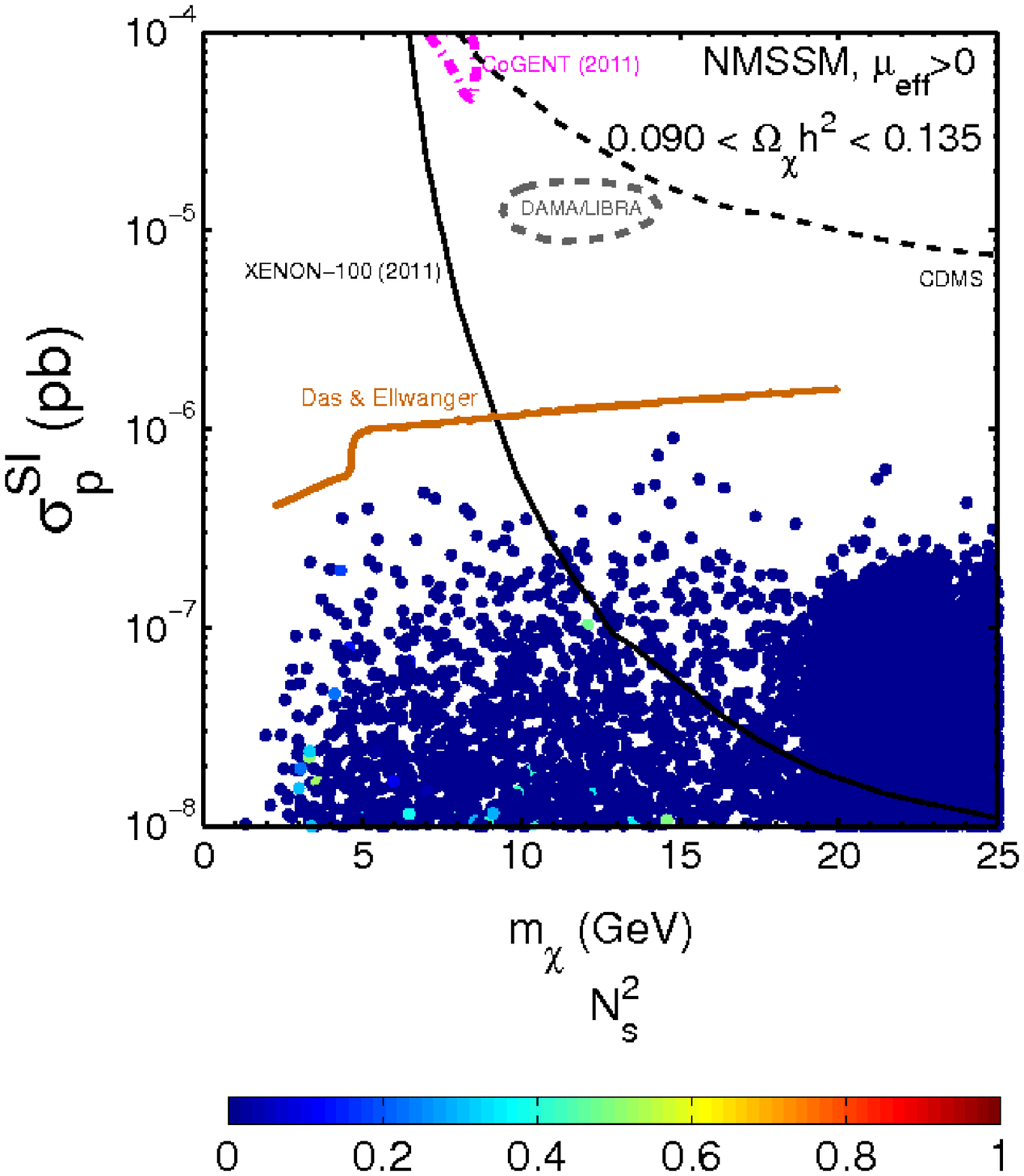}
 	\includegraphics[width=0.461\linewidth, keepaspectratio]{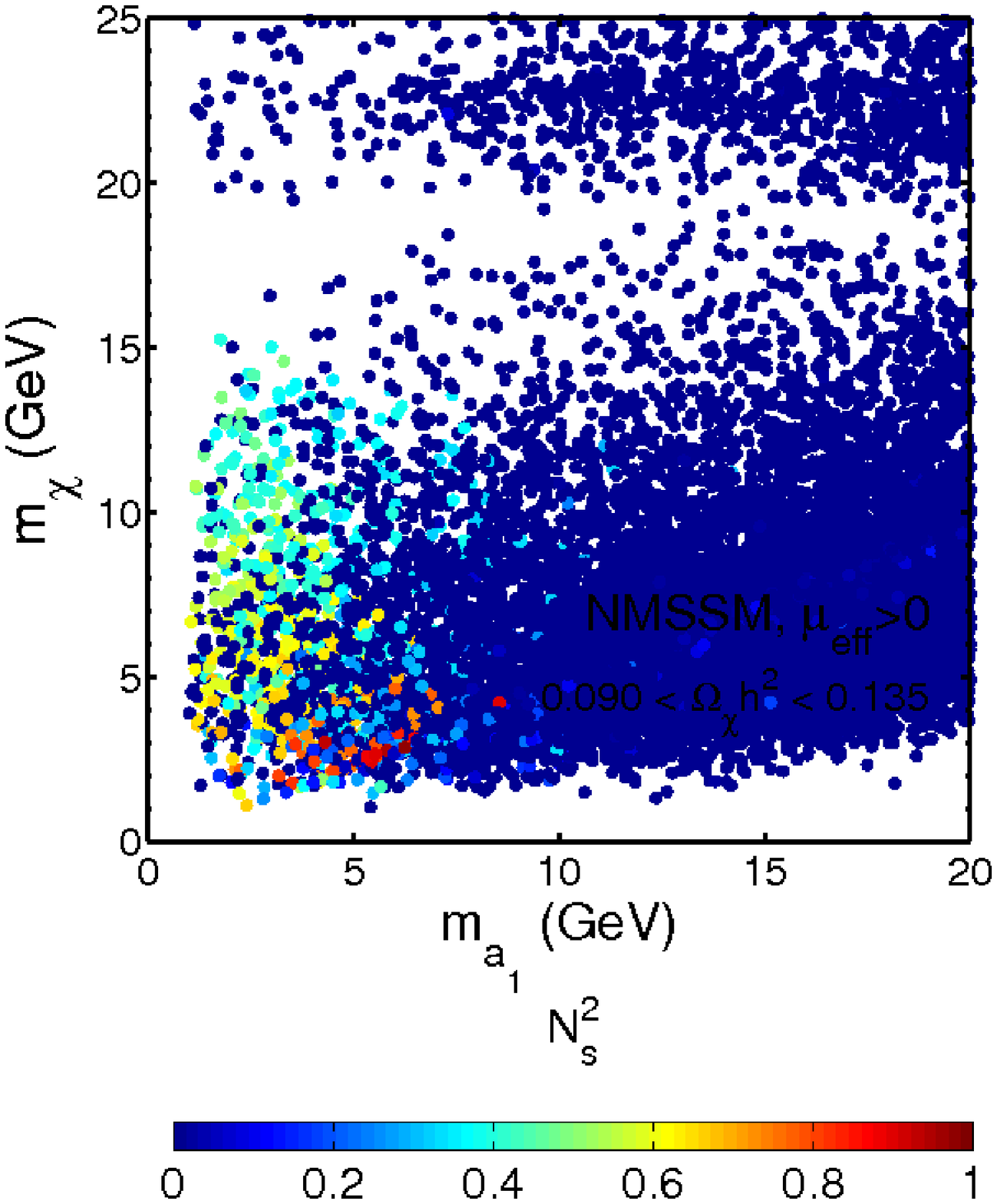}
      \end{center}
      \caption{{\it Left panel:} $\mchi$ \vs\ $\sigsip$ in the NMSSM,
        for the points from our initial NS scan that survive
        subsequent $2\sigma$ hard cuts on $\abundchi$ and when using
        constraints from colliders (including flavour physics).  The
        colour scale indicates the singlino fraction of each
        respective neutralino LSP. The experimental bounds and
        theoretical limits are identical to those displayed in the
        right panel of Fig.\,\ref{fig:oh2mchisigmchi_nmssm}.  {\it
          Right panel:} $m_{a_1}$ \vs\ $\mchi$ in the NMSSM, using the
        same points as those displayed in the left panel.}
      \label{fig:ma1mchi_nmssm}
\end{figure}

We present the results of our scan in the left panel of
Fig.\,\ref{fig:oh2mchisigmchi_nmssm}, displayed in the
$\mchi$\,-\,$\abundchi$ plane. Cyan points
correspond to those resulting from our initial scan, where collider
constraints have been invoked in the relevant likelihood as hard cuts,
differing from our MSSM scan, as described in
Sec.\,\ref{sec:procedure}, but with the constraints on
$\abundchi$ and those from flavour physics being invoked using a
Gaussian assumption, as with our MSSM scan.  The magenta points in the
left-panel of Fig.\,\ref{fig:oh2mchisigmchi_nmssm} then correspond to
those of the cyan points that survive a subsequent
$2\sigma$ hard cut utilising flavour physics constraints.  For both
data sets, we observe that it is possible to generate points in the
NMSSM that yield a relic density within 2$\sigma$ of the experimental
central value from WMAP, which are indicated by horizontal black
dashed lines.  However, as in the MSSM case, we can see many points
where their associated relic density is many orders of magnitude
larger than $0.1$, due to premature thermal freeze-out brought about
by the insufficient annihilation rates of the respective LSP cases.

In the right panel of Fig.\,\ref{fig:oh2mchisigmchi_nmssm}, we
demonstrate the effect of applying our different hard cuts on the
$\mchi$\,-\,$\sigsip$ plane. The light green and red points
correspond to those points that survive respective $2\sigma$ hard cuts
on $\abundchi$ on the cyan and magenta points displayed in the left
panel of Fig.\,\ref{fig:oh2mchisigmchi_nmssm}.  For comparison, we
also superimpose current limits from CoGENT, DAMA/LIBRA (with ion
channelling), CDMS-II and XENON-100.  We observe that the
majority of our selected points typically generate values of the SI
elastic scattering cross section in the range
$\sigsip\lesssim10^{-7}\pb$.  We also note that many of the green
points generate much larger values of $\sigsip$, up to
$2\times10^{-5}\pb$, however, the bino fraction of
these neutralino LSPs are too small to survive our hard cuts involving
flavour physics constraints.

\begin{figure}[t]
	\begin{center}
 	\includegraphics[width=0.495\linewidth, keepaspectratio]{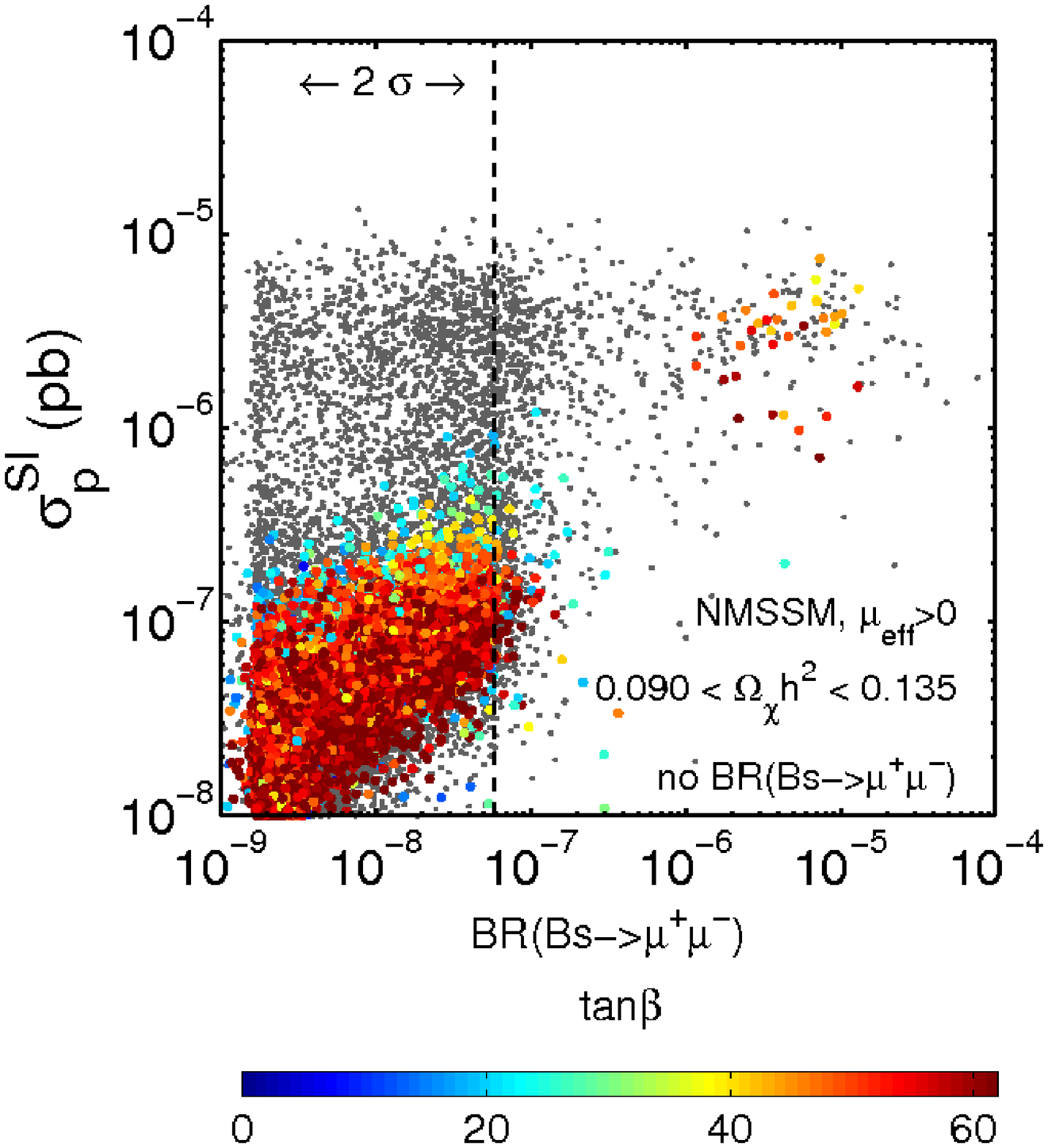}
 	\includegraphics[width=0.485\linewidth, keepaspectratio]{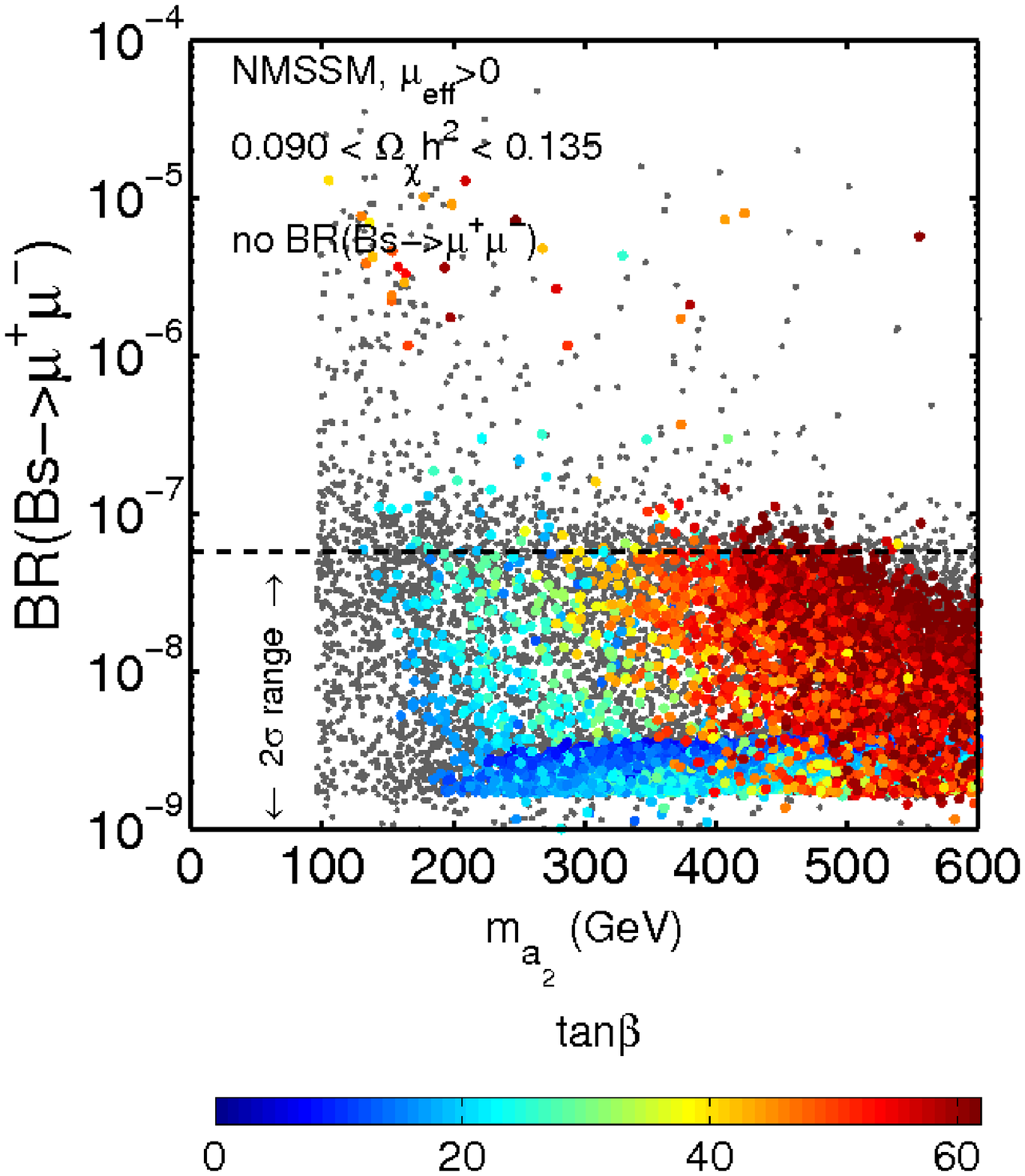}
      \end{center}
      \caption{{\it Left panel:} $\brbsmumu$ \vs\ $\sigsip$ in the
        NMSSM.  Grey points: points resulting from our initial NS
        scan that survive a subsequent $2\sigma$ hard cut on
        $\abundchi$.  Coloured points: those of the grey points
        displayed which survive further $2\sigma$ hard cuts from
        colliders (including flavour physics constraints, except that
        relating to $\brbsmumu$). The colour scale displayed indicates
        the value of $\tanb$ associated with each neutralino LSP.
        {\it Right panel:} $m_{a_2}$ \vs\ $\brbsmumu$ in the NMSSM,
        using the same points as displayed in the left panel.}
      \label{fig:sig_bsmumu_nmssm}
\end{figure}

From Fig.\,\ref{fig:oh2mchisigmchi_nmssm} (right panel) we also
observe that our results, following our $2\sigma$ hard cuts, coincide
well with the corresponding upper limits on $\sigsip$ estimated by
Das\,\&\,Ellwanger~\cite{Das:2010ww} (orange curve) using their
default model regarding the strange quark content of nucleons.
Despite this agreement, we also note that the studies conducted by
V\'{a}squez {\it et al.}~\cite{Vasquez:2010ru} and Cao {\it et al.}~\cite{Cao:2011re} 
generate results that include a small population of
neutralino LSPs with $\sigsip$ as large as $10^{-2}\pb$ with
$\mchi\sim10\gev$.  We note that the $\sigsip$ values of these points
grossly violate the DD limits from XENON-100 by up to four orders of
magnitude, as do the majority of the points we obtain possessing
$\sigsip\gtrsim10^{-6}\pb$, and hence their omission do not effect our
conclusions regarding the viable regions of NMSSM parameter space that
are consistent with such current experimental limits.  Moreover, we
also remind the reader that, as described in
Sec.\,\ref{sec:procedure}, we implicitely invoke our collider
constraints as $2\sigma$ hard cuts during our scan of the NMSSM, and
hence omit points that grossly violate such constraints.

In Fig.\,\ref{fig:mh1mchi_nmssm}, we plot the respective values of
$m_{h_1}$ against $\mchi$, associated
with the points from our initial scan surviving subsequent $2\sigma$
hard cuts using flavour physics constraints. We observe that, for
$\mchi\lesssim15\gev$, the reason why these points survive our hard
cuts on $\abundchi$ is predominantly because of the existence of an
$s$-channel resonance annihilation of the neutralino LSP via the
lightest singlet-like Higgs, i.e., where $2\mchi\simeq \mhone$.  This
can be seen in the left panel of Fig.\,\ref{fig:mh1mchi_nmssm}, where
we plot the singlet fraction, ${S_S^2}$, of the lightest
Higgs as a third axis.  The points corresponding to the LSP
annihilating through the Higgs resonance also possess the lightest
Higgs that is mostly singlet, which is essential in order to evade
collider constraints on $\mhone$.  We also observe points with $\mchi
\gsim 15\gev$ that also survive our hard cuts on $\abundchi$. These
points correspond to those configurations where one has very light
slepton masses for all three generations, by virtue of our relaxation
of the universality of slepton masses, allowing one to obtain a small
enough relic density through  additional $t$-channel
annihilations, as is the case in the MSSM.  For masses of the lightest
Higgs in the range $\mhone\gsim 114.4\gev$ we also observe the onset
of the MSSM-like scenario, where the lightest Higgs is completely
doublet-like, since we are now in the LEP-allowed region.

\begin{figure}[t]
	\begin{center}
 	\includegraphics[width=0.495\linewidth, keepaspectratio]{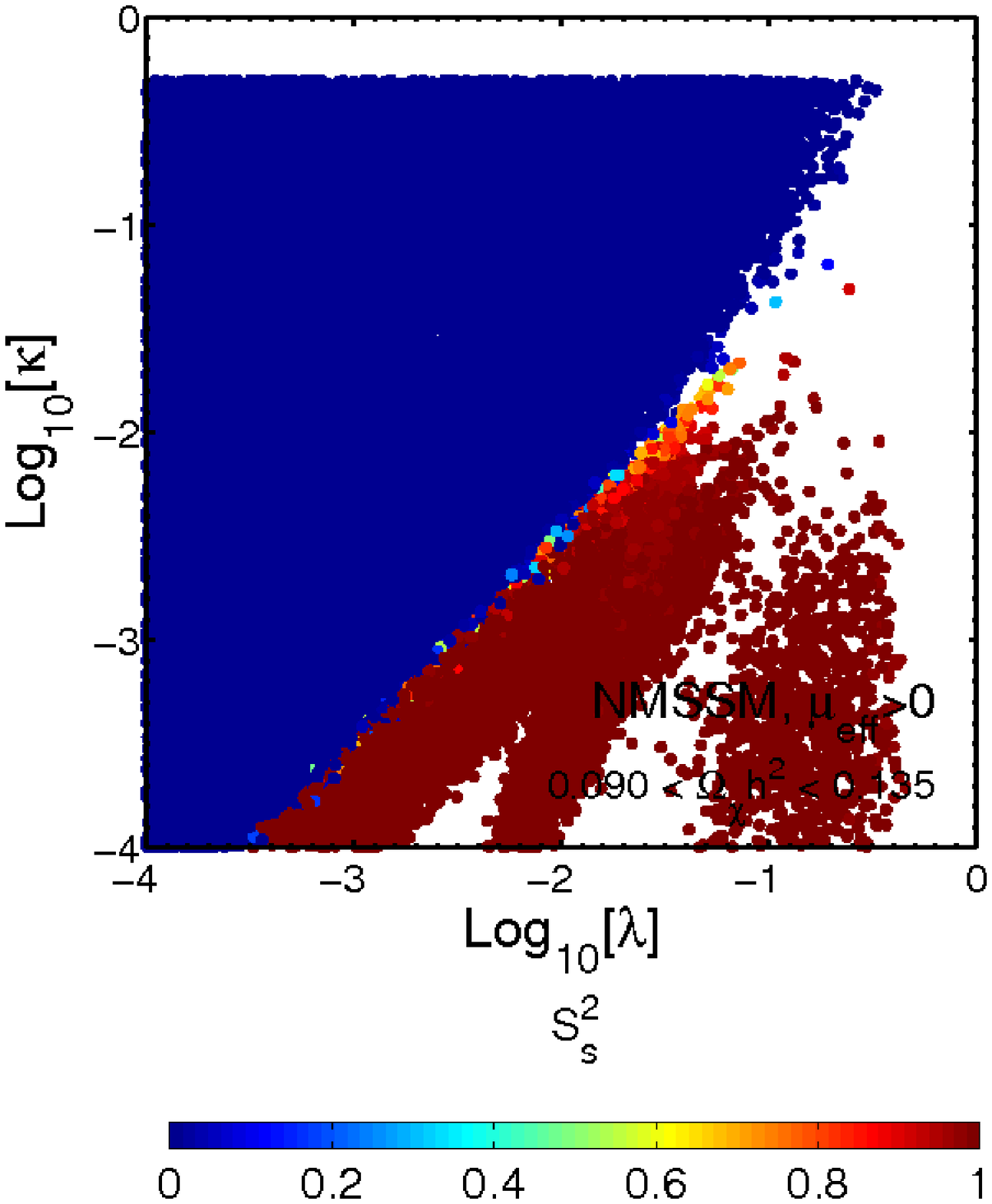}
 	\includegraphics[width=0.495\linewidth, keepaspectratio]{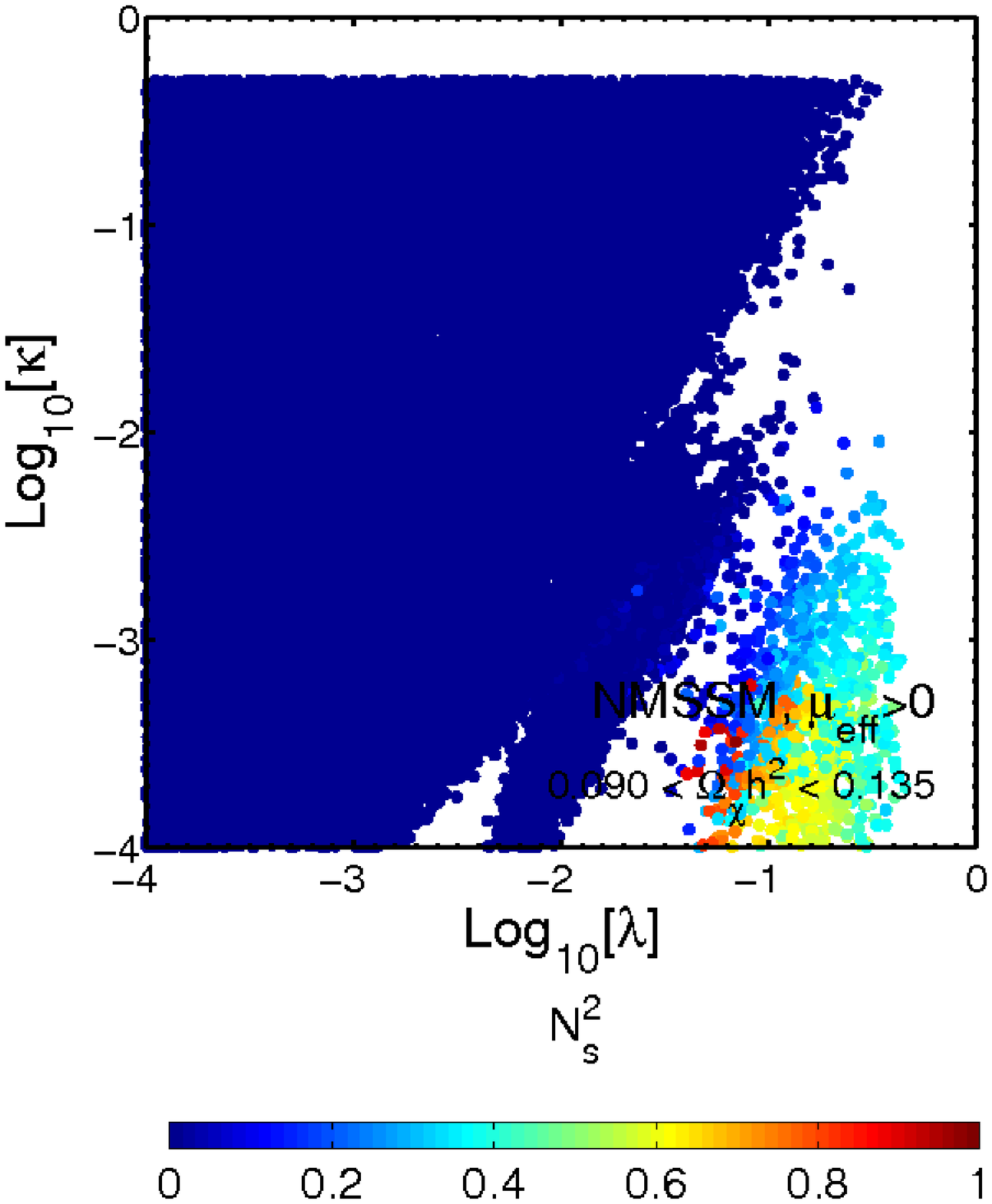}
      \end{center}
         \caption{
         {\it Left panel:} log$_{10}\,\lambda$ \vs\ log$_{10}\,\kappa$ in the NMSSM,
      for the points from our initial NS scan that survive 
      subsequent $2\sigma$ hard cuts on $\abundchi$ and
      when using constraints from colliders (including flavour physics).
      The colour scale illustrates the singlet fraction ${\rm S_{\rm S}^2}$
      of the lightest Higgs.
{\it Right panel:} Identical to the left panel except now the colour scale indicates the 
singlino fraction of each respective neutralino LSP. }
\label{fig:lambdakappa}
\end{figure}

In Fig.\,\ref{fig:ma1mchi_nmssm} we display our results in the
$\mchi$\,-\,$\sigsip$ plane (left panel) and in the
$m_{a_1}$\,-\,$\mchi$ plane (right panel).
In the left panel of
Fig.\,\ref{fig:ma1mchi_nmssm}, we can clearly see that there are only
a minority of our selected points that possess a significant singlino
component that survive our cuts and generate a value of $\sigsip
> 10^{-7}\pb$.  In the right panel of Fig.\,\ref{fig:ma1mchi_nmssm} we
observe that for some surviving configurations the lightest
pseudoscalar is lighter than the lightest neutralino. For these
configurations, if the neutralino LSP also possesses a significant
singlino component, annihilations where pseudoscalars are in the final
state become significant.  This explains the presence of those few
singlino-dominated points in the right panel of
Fig.\,\ref{fig:mh1mchi_nmssm}, where we present the singlino fraction,
$N_s^2$, of the neutralino LSP of the points displayed in the left
panel of Fig.\,\ref{fig:mh1mchi_nmssm} (via a colour scale), residing
below those undergoing the resonance annihilations via the lightest
scalar Higgs (i.e., $2\mchi\simeq\mhone$).  Alternatively, there is
the possibility of configurations possessing a neutralino LSP that
undergoes a resonance annihilation via the pseudoscalar Higgs (i.e.,
$2\mchi \simeq m_{a_1}$), which is also evident in
Fig.\,\ref{fig:ma1mchi_nmssm}. Such annihilation channels are
evidently crucial in yielding an overall annihilation rate
sufficiently large enough to generate the correct relic density
(within the $2\sigma$ range) necessary to survive the invoked hard
cuts.

The consequences of imposing the constraints from flavour physics are
much more severe.  This can be seen in
Fig.\,\ref{fig:sig_bsmumu_nmssm}, where the grey points represent
those selected by our initial NS scan that survive a subsequent
$2\sigma$ hard cut on $\abundchi$, and the coloured points represent
those of the grey points that survive further $2\sigma$ hard cuts
using flavour physics constraints except for that relating to
$\brbsmumu$. In the left panel we display these points in the
$\brbsmumu$\,-\,$\sigsip$ plane, and, in the right panel, the
$m_{a_2}$\,-\,$\brbsmumu$ plane. 
In both plots we display the value of
$\tanb$ associated with each configuration via a colour scale.

We observe that, like in the MSSM (see left panel of
Fig.\,\ref{fig:sig_bsmumu_mssm}), configurations with large
$\tan\beta$ typically yield a large SI elastic scattering cross
section within the range $10^{-6}\pb\lesssim\sigsip\lesssim10^{-5}\pb$,
but are excluded by the constraint on $\brbsmumu$ from
the Tevatron.  In the right panel, we also observe a correlation
between those points possessing large values of ${\rm tan}\,\beta$ and
small values of $m_{a_2}\lsim300\gev$ and those possessing large
values of $\brbsmumu$, analogous to the correlation between
$\brbsmumu$, $m_A$ and ${\rm tan}\,\beta$ in the MSSM displayed in
Fig.\,\ref{fig:sig_bsmumu_mssm}. However, in the NMSSM, this
correlation is evidently much weaker than in the MSSM due to the
additional degrees of freedom relating to the two pseudoscalar Higgs
masses.

Finally, in Fig.\,\ref{fig:lambdakappa} we present our results in the
$\lambda$\,-\,$\kappa$ plane with third axes displaying the singlino
composition of the lightest Higgs scalar (left panel) and singlet
composition of the lightest Higgs (right panel).  We clearly observe
that our scan points are segregated into several different regions in
the $\lambda$\,-\,$\kappa$ plane, each with a distinctive minimum
value of log$_{10}\,\lambda$, containing points that we can ascribe to one
of approximately four different categories that we will now discuss.

Starting at small values of $\lambda$ we observe the following
categories of points: (i) At $\lambda\sim10^{-4}$ we can see the onset
of the MSSM-like region, consisting of bino-dominated LSPs with
$\mchi \gtrsim 7\gev$ (see left panel of
Fig.\,\ref{fig:mh1mchi_nmssm}), with the lightest Higgs scalar being
doublet-like with $\mhone\gtrsim 114.4\gev$ in order to evade LEP
bounds.  In this case the neutralino annihilates mainly through a
$t$-channel slepton exchange as described earlier in this section.
(ii) For $\lambda\sim10^{-3}$ and $\kappa\sim10^{-4}$ we observe the
onset of a region where $\mchi\gtrsim 15\gev$ and, as with the
MSSM-like region, the LSP is bino-dominated and still annihilates
mainly via $t$-channel slepton exchange, but in this case the lightest
Higgs is singlet-like.  (iii) For $\lambda\sim10^{-2}$ we find the
onset of those points involving bino-dominated neutralino LSPs that
annihilate through the fore-mentioned $s$-channel resonance involving
a singlet-like Higgs, possessing a mass following the relation:
$\mchi\approx 2\mhone $.  iv) Lastly, for $\lambda\gtrsim 10^{-1}$ we
find neutralinos with $\mchi\lesssim 15\gev$ which are now a mixture
of both singlino and bino and, as with (iii), annihilate via a
resonance involving a singlet-like Higgs.  
As explained in Sec.\,\ref{subsec:NMSSM_results}, a degree of fine-tuning in
$A_{\lambda}$, via Eq.\,(\ref{eq:NMSSM_A_lambda}), or in $A_\kappa$,
via Eq.(\ref{eq:NMSSM_m_pseudoscalar}), is necessary
in order to generate very light singlet-like Higgses for large values
of $\lambda\sim1$.  
Such solutions can potentially generate large values of the SI elastic 
scattering cross section that are several orders of magnitude larger than 
that associated with the bino-dominated case, for $\lambda \sim 10^{-2}$, 
depending on the degree of fine-tuning involved. However, as remarked earlier 
in this section, such points are clearly excluded by the DD limits, 
particularly those from XENON-100.

\section{Summary}
\label{sec:summary}

In this paper we have re-examined the low mass neutralino region of
both MSSM and NMSSM parameter space in light of the recent
re-confirmation of a low energy excess of events observed by CoGENT,
and the recent limits arising from the non-observations by XENON-100
that are in contention with the CoGENT data. To conduct our
investigation we utilised focused NS scans when imposing a variety
of experimental constraints, associated with (i) the cosmological DM
relic abundance, (ii) direct collider search limits and (ii) collider
physics relating to flavour physics and $\deltagmtwo$, in the likelihood. Unlike the
majority of previous studies we then re-invoked each of these
constraints on the resulting scan data, this time, via a successive
series of $2\sigma$ hard cuts.

Firstly, regarding the MSSM, we relaxed the unification of gaugino
masses in order to efficiently generate light, bino-like neutralino
LSPs without violating collider constraints involving the invisible $Z$
width and the chargino mass.  We observed that for many configurations
from our initial scan producing the neutralino LSP mass $\mchi\gsim5
\gev$ the corresponding values of $\abundchi$ were small enough to be
within the WMAP $2\sigma$ range. This could be achieved predominantly
through efficient $t$-channel annihilations via stau exchange, which
are enhanced for points possessing large $\tanb$.  However, owing to
LEP constraints on the slepton masses, when we invoked collider
constraints as $2\sigma$ hard cuts, only bino-like LSPs with
$\mchi\gsim11 \gev$ survived.  Moreover, when invoking the constraints
from flavour physics, the minimum LSP mass was further increased to
approximately $\mchi\gsim13 \gev$, which is slightly heavier than the
values within the favoured CoGENT region, and much lighter than the
$28\gev$ limit claimed by Vasquez {\it et al.}~\cite{Vasquez:2010ru}.
We find that for these points it is extremely difficult to generate
values of the SI elastic scattering cross section much larger than
$10^{-5}\pb$, owing primarily to the constraints on the $\chi Hp$
coupling, which is enhanced for large $\tanb$, coming from flavour
physics, particularly the limit on $\brbsmumu$.  We emphasise that
this limit is particularly stringent within the framework of minimal
flavour violation which we have adopted but can be fairly easily
relaxed when one abandons it; see~\cite{Foster:2005wb, Foster:2006ze}. However, we
point out that solutions possessing such large values of $\sigsip$,
within our permitted mass range of $\mchi\gsim13 \gev$, are also
clearly excluded by the DD limits from XENON-100.

Regarding the NMSSM, once again, we conducted a focused NS scan in
the low LSP mass region, where (non-flavour) collider constraints were
invoked in the likelihood this time as $2\sigma$ hard cuts.
Once again, in order to efficiently generate light neutralino LSPs
without violating LEP bounds on the invisible $Z$ width and the
chargino mass, gaugino unification was relaxed. Whilst the majority of
points surviving collider constraints were found to be bino-like, we
found that a significant proportion of these points also evaded
collider constraints by adopting a partial singlino component, made
possible through the extended Higgs sector of the NMSSM. In such
cases, we found that for light LSPs possessing $\mchi\lsim15 \gev$ we
were able to generate values of $\abundchi$ small enough to lie within
the WMAP $2\sigma$ range through $s$-channel resonance annihilations
via the lightest CP-even Higgs.  Consequently, in order to evade LEP bounds,
in such cases the lightest Higgs had to be singlet-like, very much
decoupled from the doublet Higgs.  
This could be achieved by adopting the limit $\lambda\rightarrow0$,
in which case bino-dominated LSPs are favoured, or by finely-tuning
$A_\lambda$, generally resulting in a combination of bino-singlino
LSP configurations and a hierarchy between $\lambda$ and $\kappa$
where $\kappa\ll\lambda$ for $\lambda\gg0$ that we explicitly described,
or by finely-tuning $A_\kappa$.
However, for $\mchi\gsim15 \gev$, we
found that points can generate $\abundchi$ small enough to survive our
cuts by virtue of additional $t$-channel annihilations via very light
sleptons, possible through our relaxation of the universality of
slepton masses, or in some cases through $s$-channel annihilations via
the pseudoscalar Higgs.  We found that the points from our initial
scan generated values of $\sigsip$ up to approximately
$2\times10^{-5}\pb$. However, we found that many of these points
possessed too large a bino fraction to survive the flavour physics
constraints, particularly that coming from $\brbsmumu$, that we later
re-invoked as $2\sigma$ hard cuts, finding results very similar to those
obtained in~\cite{Das:2010ww} where $\sigsip \lsim 10^{-6}\pb$. We
also highlight that many of these points with $\mchi\gsim10\gev$ are
excluded by DD limits from XENON-100, like many of the points
discussed by other authors, including~\cite{Vasquez:2010ru,Cao:2011re}, 
that possess $\sigsip$ as large as
$10^{-2}\pb$, but are generated without invoking experimental
constraints as $2\sigma$ hard cuts, as we do, in order to omit points
that grossly violate individual constraints.

\acknowledgments{\noindent We would like to thank J.\,I.\,Collar and
  V.~ Kudryavtsev for their helpful comments regarding the CoGENT
  experiment and data analysis, and also to T. Varley for his participation 
  early in the project. 
  DTC is supported by the Science Technology and Facilities Council. 
  DEL-F is supported by the French ANR TAPDMS ANR-09-JCJC-0146 
  and would like to thank the Science Technology and Facilities Council 
  for its support at an earlier stage of this project. 
  LR is funded in part by the Foundation for Polish Science and by the 
  EC 6th Framework Programme MRTN-CT-2006-035505. 
  R. RdA would like to thank the support of the Spanish MICINN's
  Consolider-Ingenio 2010 Programme under the grant MULTIDARK 
  CSD2209-00064.}



\end{document}